\documentclass[%aip,
 amsmath,amssymb,
reprint,%
]{revtex4-1}
\usepackage{color}
\usepackage{graphicx}% Include figure files
\usepackage{dcolumn}% Align table columns on decimal point
\usepackage{bm}% bold math
\usepackage[utf8]{inputenc}
\usepackage[T1]{fontenc}
\usepackage{mathptmx}

\begin{document}

\preprint{}
\title{A coarse-grained polymer model for studying the glass transition}
\author{Hsiao-Ping Hsu}
\email[]{hsu@mpip-mainz.mpg.de}
\affiliation{Max-Planck-Institut f\"ur Polymerforschung, Ackermannweg 10, 55128, Mainz, Germany}
\author{Kurt Kremer}
\email[]{kremer@mpip-mainz.mpg.de}
\affiliation{Max-Planck-Institut f\"ur Polymerforschung, Ackermannweg 10, 55128, Mainz, Germany}

\begin{abstract}
   To study the cooling behavior and the glass transition of polymer melts
in bulk and with free surfaces a coarse-grained weakly semi-flexible polymer 
model is developed. Based on a standard bead spring model with purely repulsive 
interactions an attractive potential between non-bonded monomers is added,
such that the pressure of polymer melts is tuned to zero. Additionally, the commonly used 
bond bending potential
controlling the chain stiffness is replaced by a new bond bending potential. 
For this model, we show that the Kuhn length and 
the internal distances along the chains in the melt only very weakly depend on 
temperature, just as for typical experimental systems. 
The glass transition is observed by the temperature dependency of the melt density
and the characteristic non-Arrhenius slowing down of the chain mobility. 
The new model is set to allow for a fast switch between models, for which a wealth of 
data already exists. 
     
\end{abstract}

%\keywords{}

\maketitle

%bending potential
%prevent chains stretching out at low temperature

  Polymer materials are omnipresent in our daily life with applications 
in medicine, technology as well as as `simple' commodities to name a few.
Very often these materials are in the glassy 
state~\cite{Hrushikesh2017}.  
In the liquid more rubbery state the viscosity dramatically 
increases close to the glass transition temperature $T_g$ in a non-Arrhenius 
way~\cite{Boyd1994,Angell1995,Paluch2001,Berthier2011}. 
This slowing down of the chain mobility is of both high scientific 
and technological interest. 
Experimentally, $T_g$ of polymers can be determined as such by observing the
change in the heat capacity of polymers using 
differential scanning calorimetry ({DSC)}~\cite{Mathot1994}, or by measuring the 
thermal expansion coefficient using thermo mechanical analysis (TMA)~\cite{Bird1987}.
However, the nature of the glass transition is still not fully
understood~\cite{Angell1988,Angell1991,Binder2005, Barrat2010,Stillinger2013,Ediger2014}.
It is the purpose of this communication to present a most simple, efficient bead spring model, which
allows to study these effects and which can make contact to the huge body of simulation
data available in the literature. 

  Computer simulations play an important role in investigating
the structure and molecular motion (viscosity) of polymeric systems under 
a variety of different conditions.
For studying glassy polymers, both atomistic and coarse-grained models
are widely used in the literature~\cite{Binder2005, Barrat2010}. 
The structure and thermal behavior of fluid mixtures can also be analyzed by tuning relative resolution 
in a recently developed hybrid model combing the fine-grained and coarse-grained models~\cite{Chaimovich2015}.
Our aim is to eventually study generic properties of  
large and highly entangled polymer melts in bulk, in confinement and with 
free surfaces as a function of temperature within accessible computing times. 
For this  we adopt a highly efficient coarse-grained model~\cite{Kremer1990}.
Usually in these models the excluded volume interaction is taken care of by a purely 
repulsive Lennard-Jones (LJ) potential, 
the Weeks-Chandler-Andersen (WCA) potential~\cite{Kremer1990}, which prevents the study 
of surfaces~\cite{Duenweg1997,Kopf1997} and displays a rather high pressure
($P \approx 5.0\epsilon/\sigma^3$, $T=1.0 \epsilon/k_B$, density $\rho = 0.85 \sigma^{-3}$ 
in standard Lennard-Jones (LJ) units of energy and length, and $k_B$ being the Boltzmann factor). 
{To reduce the pressure the cut-off of the WCA potential for
non-bonded pairs of monomers is often doubled from $r_{\rm cut}=2^{1/6}\sigma$ to
$r_c=2r_{\rm cut}$, resulting $P=1.0\epsilon/\sigma^3$~\cite{Bennemann1998,Binder1999,Buchholz2002,Binder2003,Schnell2011,Frey2015}.}
The two main shortages of this setting are: (1)
There is a small discontinuity in the force at the cut-off making microcanonical runs
impossible and 
(2) the pressure is still not very close to zero. Furthermore, chain stiffness usually
is taken into account by a bond bending potential~\cite{Everaers2004,Svaneborg2018a,Svaneborg2018b}, 
which tends to stretch the chains out with decreasing 
temperatures~\cite{Grest2016}. As will be shown below, this leads to rather artificial
chain conformations upon cooling, while in experiment chain conformations only very
weakly depend on temperature~\cite{Wind2003,Fetters2007}. Our new coarse-grained model is set to overcome these shortages.

Our starting point is the standard bead spring model
(BSM)~\cite{Kremer1990} with a weak bending
elasticity~\cite{Everaers2004}
(the bending strength $k_\theta=1.5\epsilon$) for which
a huge body of data already
exists (see e.g.~\cite{Zhang2014,Moreira2015,Hsu2016,Hsu2018a,Hsu2018b,Svaneborg2018a}). While focusing on $k_\theta=1.5\epsilon$, our approach easily applies to other bending constants as well. At the standard melt density of $0.85 \sigma^{-3}$ ($\sigma$ being the unit of
length) the weak bending elasticity combined with the chain packing
result in an entanglement length of only $N_e=28$ monomers.
$N_e=28$ is small enough to allow for extremely efficient simulations
of highly entangled, huge polymeric systems,
while at the same time the subchain of length $N_e$ is already well described
by a Gaussian chain. The purpose of this communication is to
replace/extend the WCA excluded volume interaction potential to
arrive at a pressure of $P=0.0 \epsilon/\sigma^3$, which allows
to study {free surfaces in interaction with gases, liquids, and particles
for example}, and to replace the standard bending potential
$U_{\rm BEND}^{\rm (old)}(\theta)=k_\theta(1-\cos \theta)$ by a new
modified $U_{\rm BEND}(\theta)$, which should lead to the typical very
weak temperature dependence of chain conformations in melts. The close resemblance to the
standard semiflexible bead spring model will allow to switch
``on the fly'' between the models and to make use of the already broadly
available data.

\begin{figure*}[t!]
\begin{center}
(a)\includegraphics[width=0.30\textwidth,angle=270]{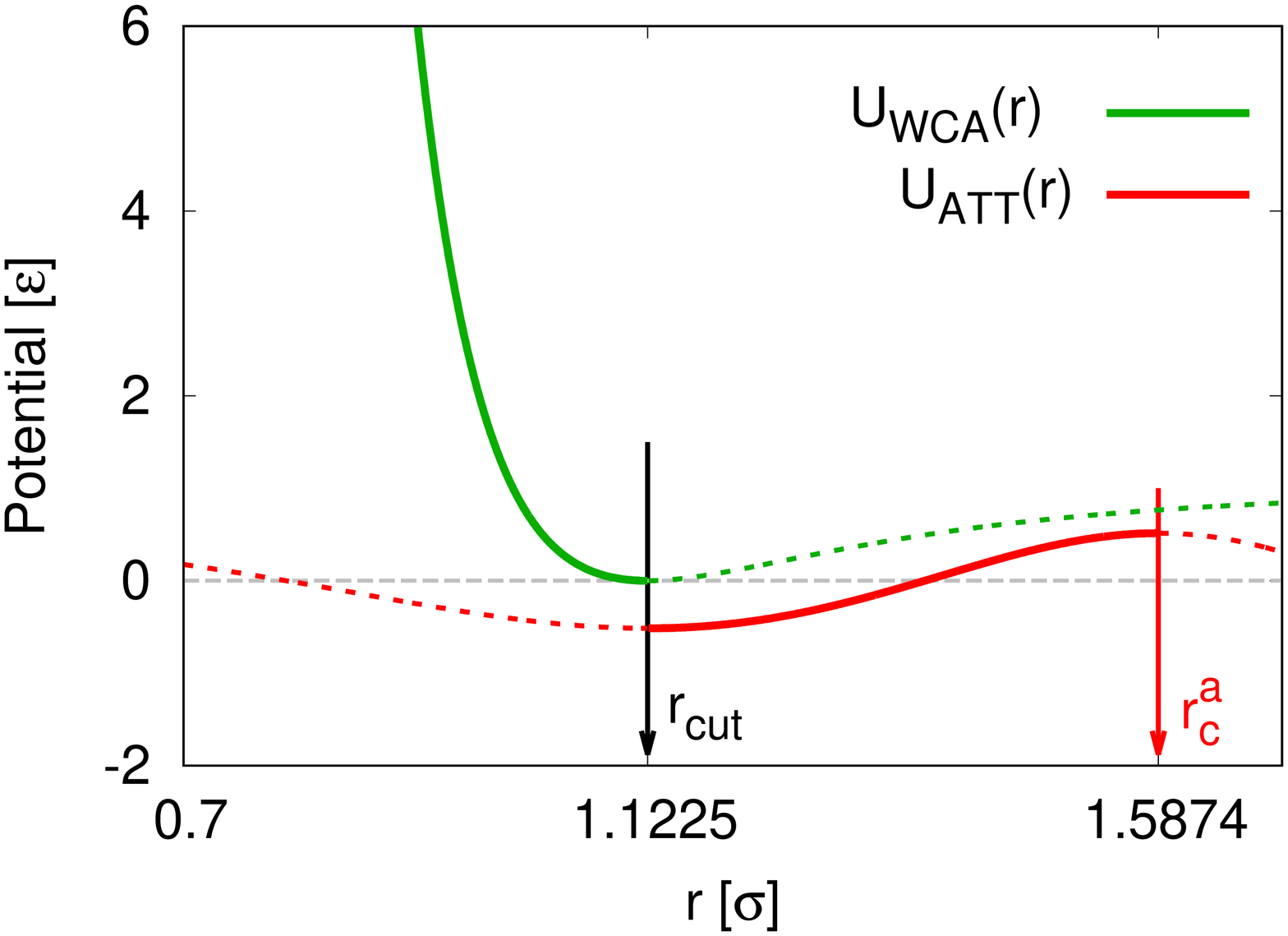} \hspace{0.1truecm}
(b)\includegraphics[width=0.30\textwidth,angle=270]{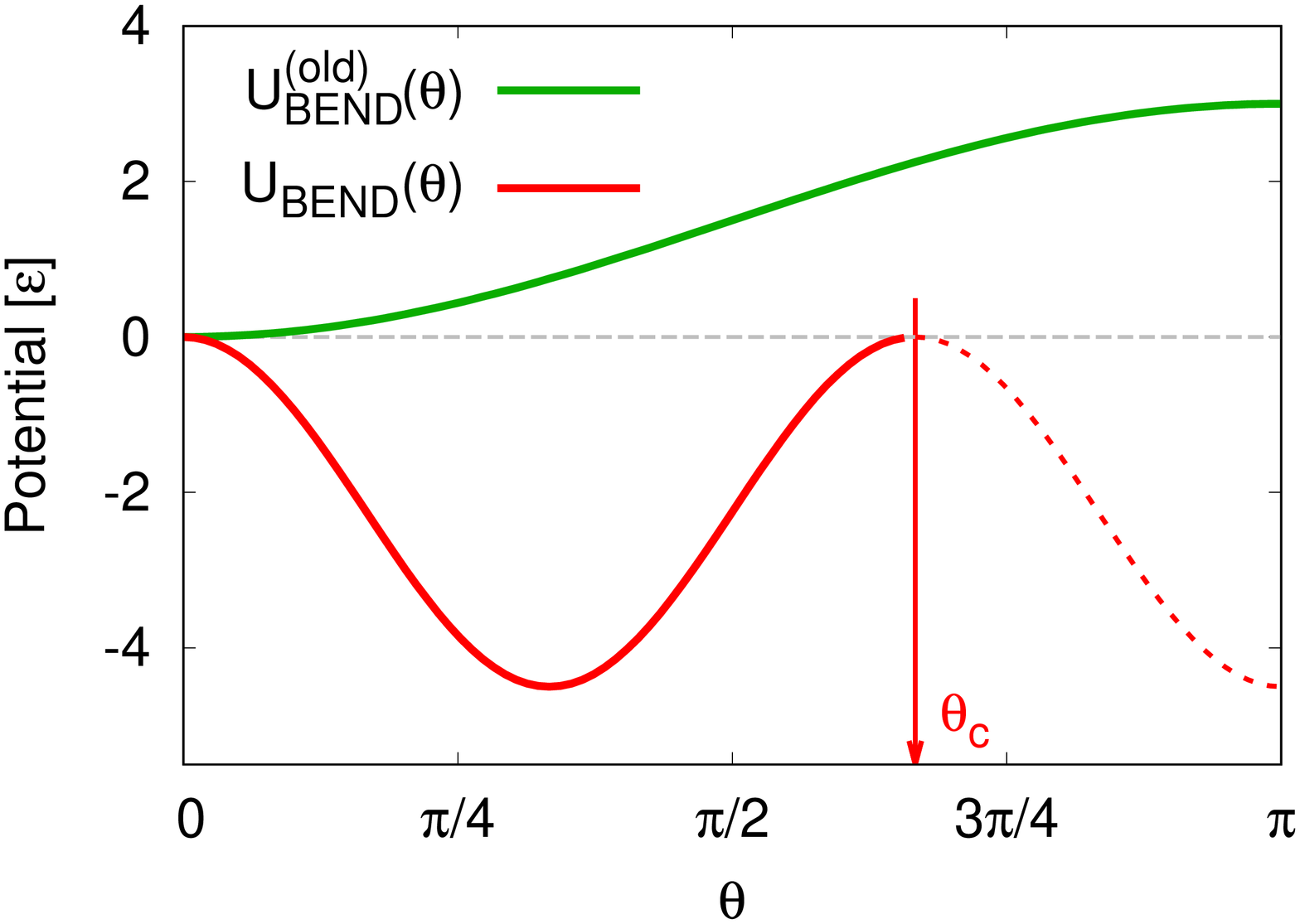}
\caption{(a) Non-bonded and short-range repulsive potential $U_{\rm WCA}(r)$ and
attractive potential $U_{\rm ATT}(r)$
with $\alpha=0.5145\epsilon$ \{Eq.~(\ref{eq-Uatt})\} plotted as a function of distance $r$.
(b) Standard and new bond bending potentials, $U^{\rm (old)}_{\rm BEND}(\theta)$
with $k_\theta=1.5\epsilon$ and $U_{\rm BEND}(\theta)$ with
$a_\theta=4.5\epsilon$, $b_\theta=1.5$ \{Eq.~(\ref{eq-Ubend})\}, 
%$\theta_c=(3\pi/2)/b_\theta$, 
plotted as a function of bond angle $\theta$.
In (a)(b), the cut-off values are pointed by arrows.}
\label{fig-U}
\end{center}
\end{figure*}

\begin{figure*}[thb!]
\begin{center}
(a)\includegraphics[width=0.30\textwidth,angle=270]{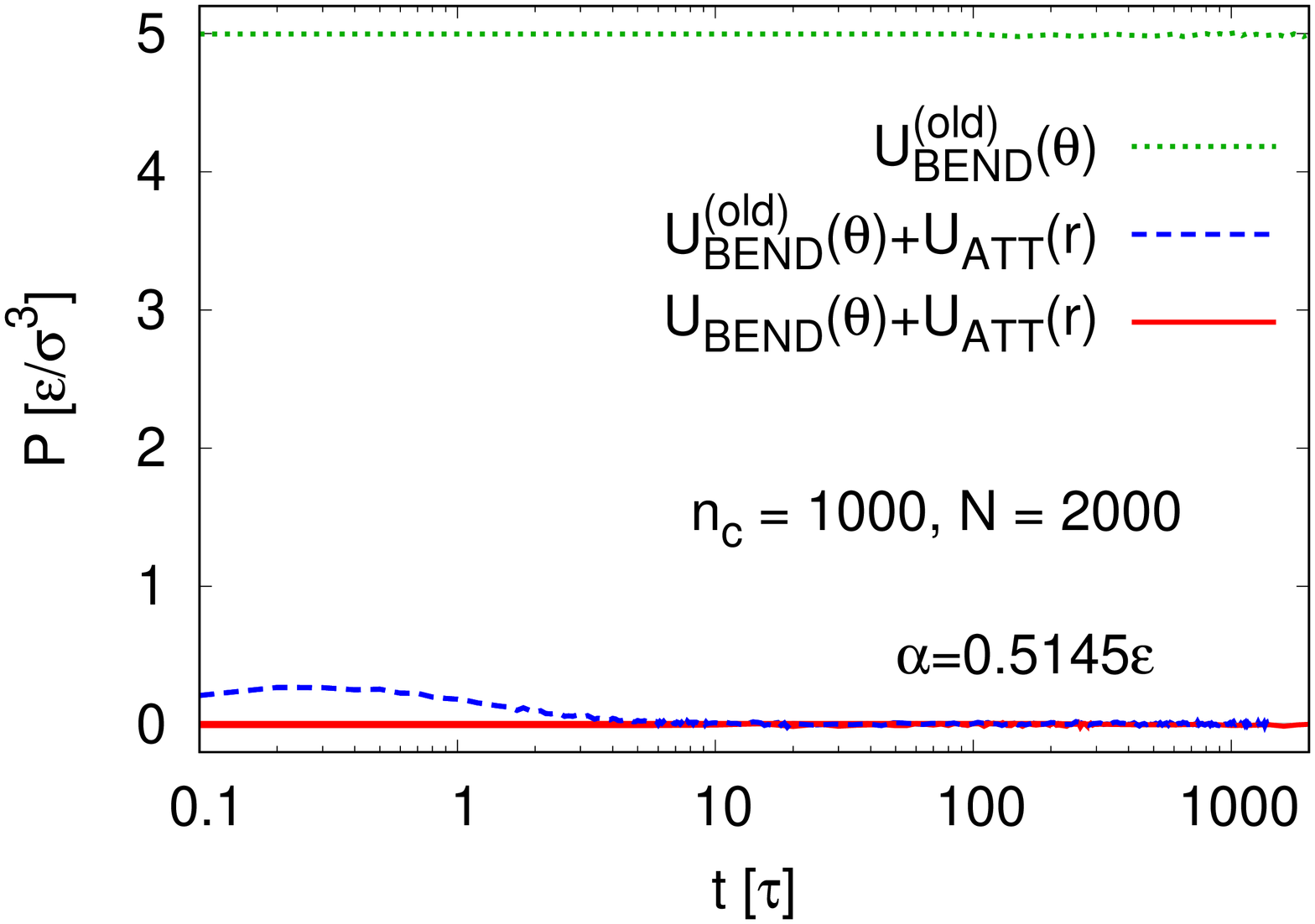} \hspace{1.2truecm}
(b)\includegraphics[width=0.30\textwidth,angle=270]{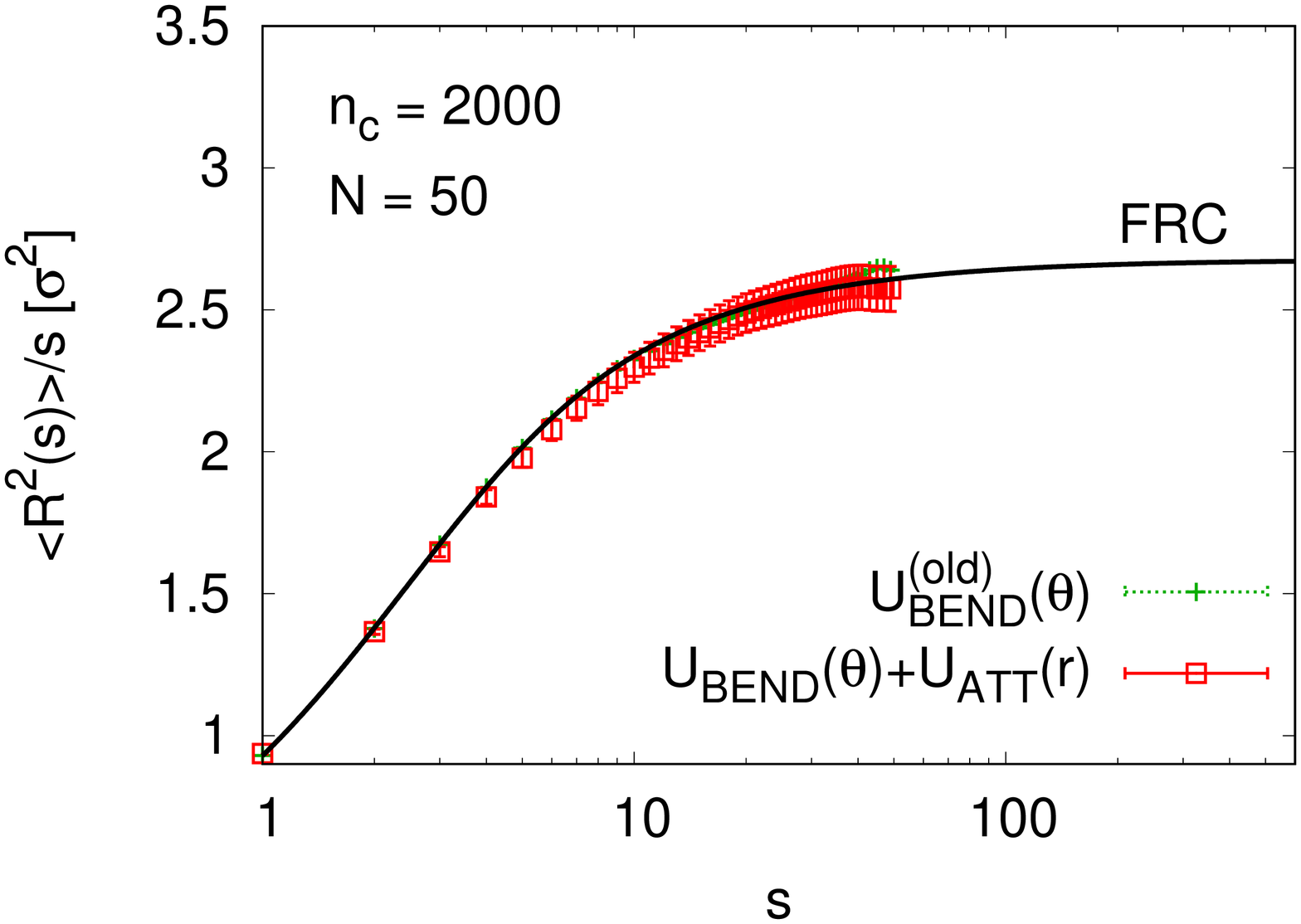}\\
(c)\includegraphics[width=0.30\textwidth,angle=270]{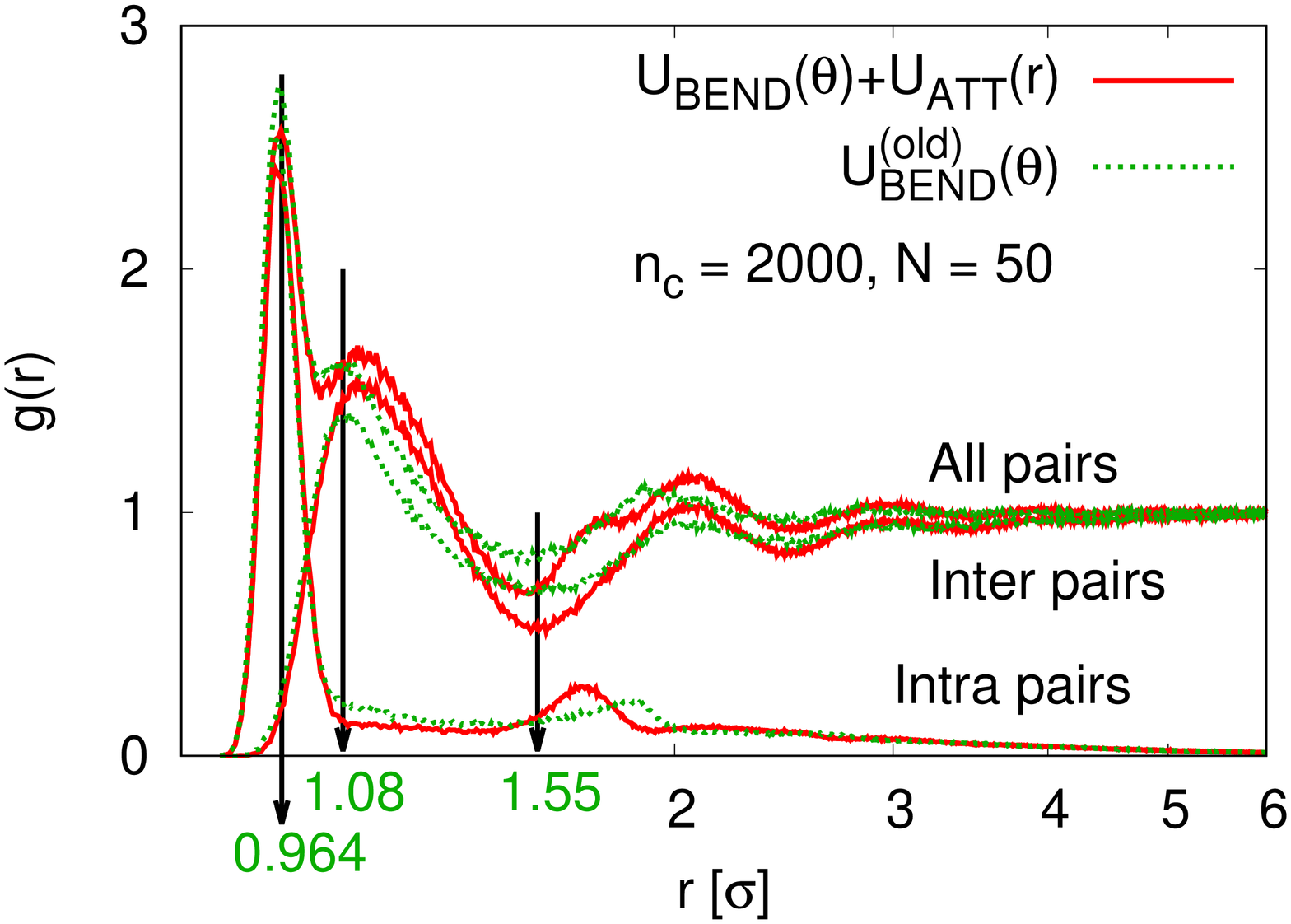} \hspace{1.2truecm}
(d)\includegraphics[width=0.30\textwidth,angle=270]{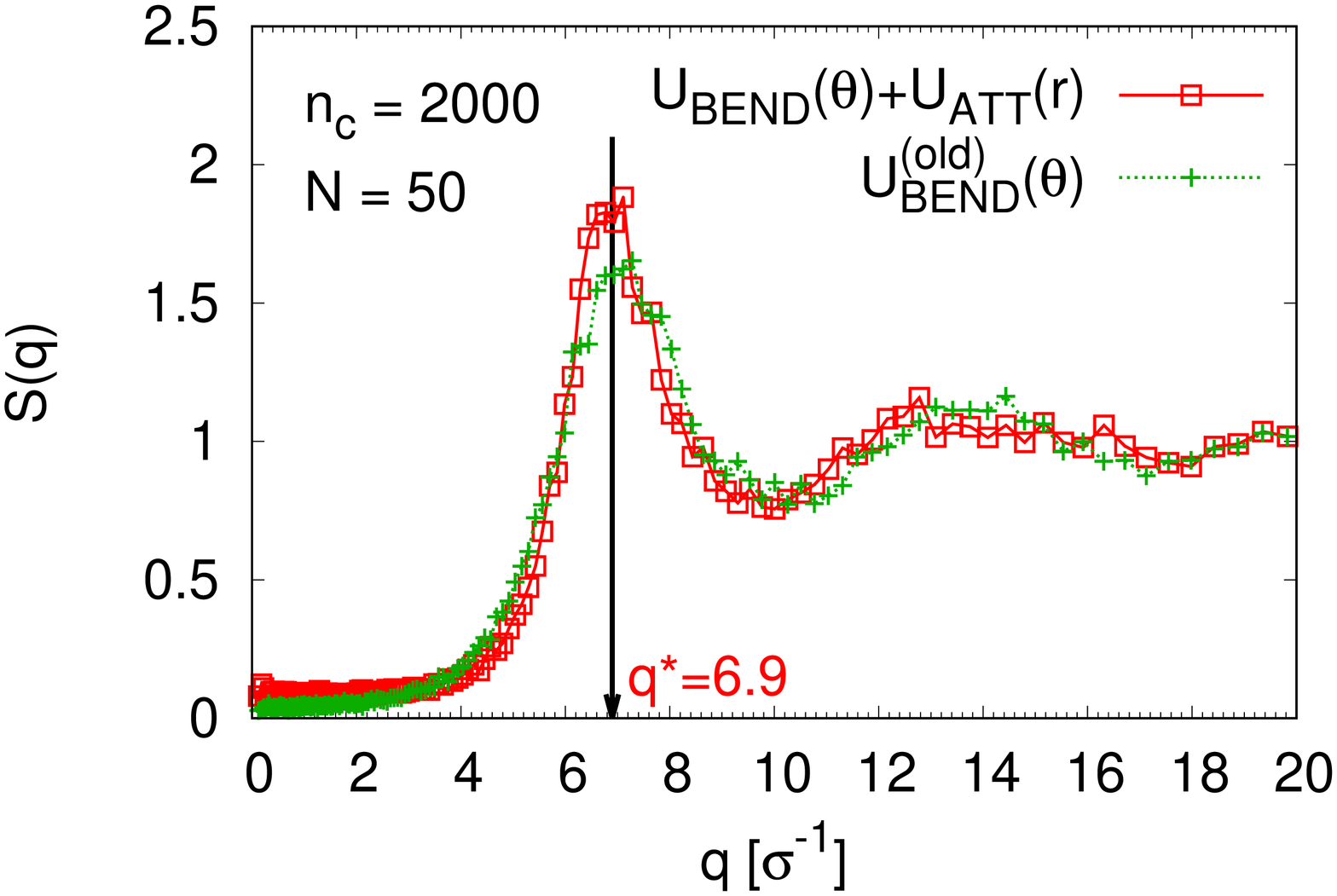}\\
\caption{(a) Pressure $P$ plotted versus the relaxation time $t$.
(b) Rescaled mean square internal distance, $\langle R^2(s) \rangle/s$, plotted
versus the chemical distance $s$ between two monomers along the same chain.
(c) Radial distribution function $g(r)$ plotted as a function of $r$ for all,
inter, and intra pairs of monomers, as indicated.
(d) Collective structure factor $S(q)$ plotted versus the wave factor $q$.
Polymer melts at $T=1.0\epsilon/k_B$ described by the standard BSM with additional potentials
$U^{\rm (old)}_{\rm BEND}(\theta)$, $U_{\rm ATT}(r)$, and $U_{\rm BEND}(\theta)$ are shown, as indicated.}
\label{fig-conf}
\end{center}
\end{figure*}

In a first step we add an attractive well to the WCA excluded volume 
in order to reduce the pressure in the system from $P=5.0\epsilon/\sigma^3$
to $P=0.0\epsilon/\sigma^3$. For this we add $U_{\rm ATT}(r)$ 
(see Figure~\ref{fig-U}a),
\begin{eqnarray}
U_{\rm ATT}(r)=\left\{\begin{array}{ll}
\alpha \left[ \cos(\pi \left(\frac{r}{r_{\rm cut}} \right)^2) \right]
   ,& {r_{\rm cut}} \leq r < r^a_{c} \\
& \\
0  ,& {\rm otherwise}
\end{array} \right . \,, 
\label{eq-Uatt}
\end{eqnarray}
between all non-bonded monomers.  $U_{\rm ATT}(r)$ is set to not alter the 
local bead packing. 
It is chosen to have zero force at the cut-off as well as at the
contact point between the two parts of the potential at $r_c=2^{1/6}\sigma$,
which is needed in the case microcanonical simulations are performed.
As illustrated in Figure~\ref{fig-conf}a, adding this term to the standard model
equilibrates and reduces the pressure to zero in less than $5\tau$
($\tau$ being the standard LJ unit of time). 
This time corresponds to a small, local bead displacement of about $1\sigma$,
for which the 
characteristic time is~\cite{Hsu2016} $\tau_0 \approx 2.89\tau$. 
Furthermore, since the number of particles $Z$ in the interaction range 
$r_c^a = 1.5874\sigma$ is $ \approx 15$ instead of $ \approx 45$ 
at $r_c = 2.25\sigma$ $(P=1.0\epsilon/\sigma^3)$ or $\approx 60$ 
at $r_c=2.5\sigma$ $(P=0.0\epsilon/\sigma^3)$
using the standard LJ potential, the present model is computationally significantly more efficient.
In the next step we replace the standard bond bending potential
$U^{\rm (old)}_{\rm BEND}(\theta) = k_{\theta} (1-\cos \theta)$ which would lead to a rod-like chain
in the ground state at $T=0.0\epsilon/k_B$
by a new bending potential $U_{\rm BEND}(\theta)$ with the goal to (1) 
match the chain conformations at $T=1.0 \epsilon/k_B$ and (2) to approximately
preserve them upon cooling.
Thus it should satisfy the condition that the mean square end-to-end
distance of chains, $\langle R_e^2 \rangle$, does not (preferably) or only
very weakly depend on the temperature $T$.
The new bond bending potential $U_{\rm BEND}(\theta)$ (see Figure~\ref{fig-U}b)
is chosen as
\begin{equation}
   U_{\rm BEND}(\theta) = -a_\theta \sin^2 (b_\theta \theta) 
\,, \, \, 0 < \theta <\theta_c
\label{eq-Ubend}
\end{equation}
with the bond angle $\theta$ defined by
$
   \theta=\cos^{-1}\left(  \frac{\vec{b}_j \cdot \vec{b}_{j+1}}
{\mid \vec{b}_j \mid \mid \vec{b}_{j+1} \mid} \right)
$
where $\vec{b}_j=\vec{r}_j-\vec{r}_{j-1}$ is the bond vector between
monomers $j$ and $(j-1)$ along the chain.
The fitting parameters
$a_\theta$ and $b_\theta$, and the cut-off $\theta_c = \pi/{b_{\theta}}$ where the force
$\mid \vec{F}(\theta=\theta_c)\mid=0$ are adjusted such that the estimates of
the mean square internal distance $\langle R^2(s) \rangle$ for 
all chemical distance $s$ between two monomers along the same chain follow the same curve as obtained
from the model using $U^{\rm (old)}_{\rm BEND}(\theta)$ with $k_\theta=1.5\epsilon$.
Comparing to the reference data for a polymer melt of $n_c=2000$, $N=50$
shown in Figure~\ref{fig-conf}b, we find that $a_\theta=4.5\epsilon$,
$b_\theta=1.5$. 
leads to an almost perfect match of the two systems.
Our data are also in perfect
agreement with the theoretical prediction described by a freely rotating chain
(FRC) model~\cite{Rubinstein2003,Hsu2016}.

Compared to the original model, the profiles of the
pair distribution function $g(r)$ of all, inter, and intra pairs of monomers
for polymer melts show that the two potentials $U_{\rm BEND}(\theta)$ and
$U_{\rm ATT}(r)$ only have very small effects on the local packing of monomers
(Figure~\ref{fig-conf}c).
Results of the collective structure factor $S(q)$ also show that using the
new model, the occurrence of the first peak remains at $q=q*\approx 6.9\sigma^{-1}$
indicating the same mean distance between monomers in the first neighbor shell of the polymer melt. The peak
itself is slightly higher, {indicating a slightly more structured local environment,}
in agreement with the observed weakly enhanced
bead friction.

\begin{figure*}[t!]
\begin{center}
(a)\includegraphics[width=0.30\textwidth,angle=270]{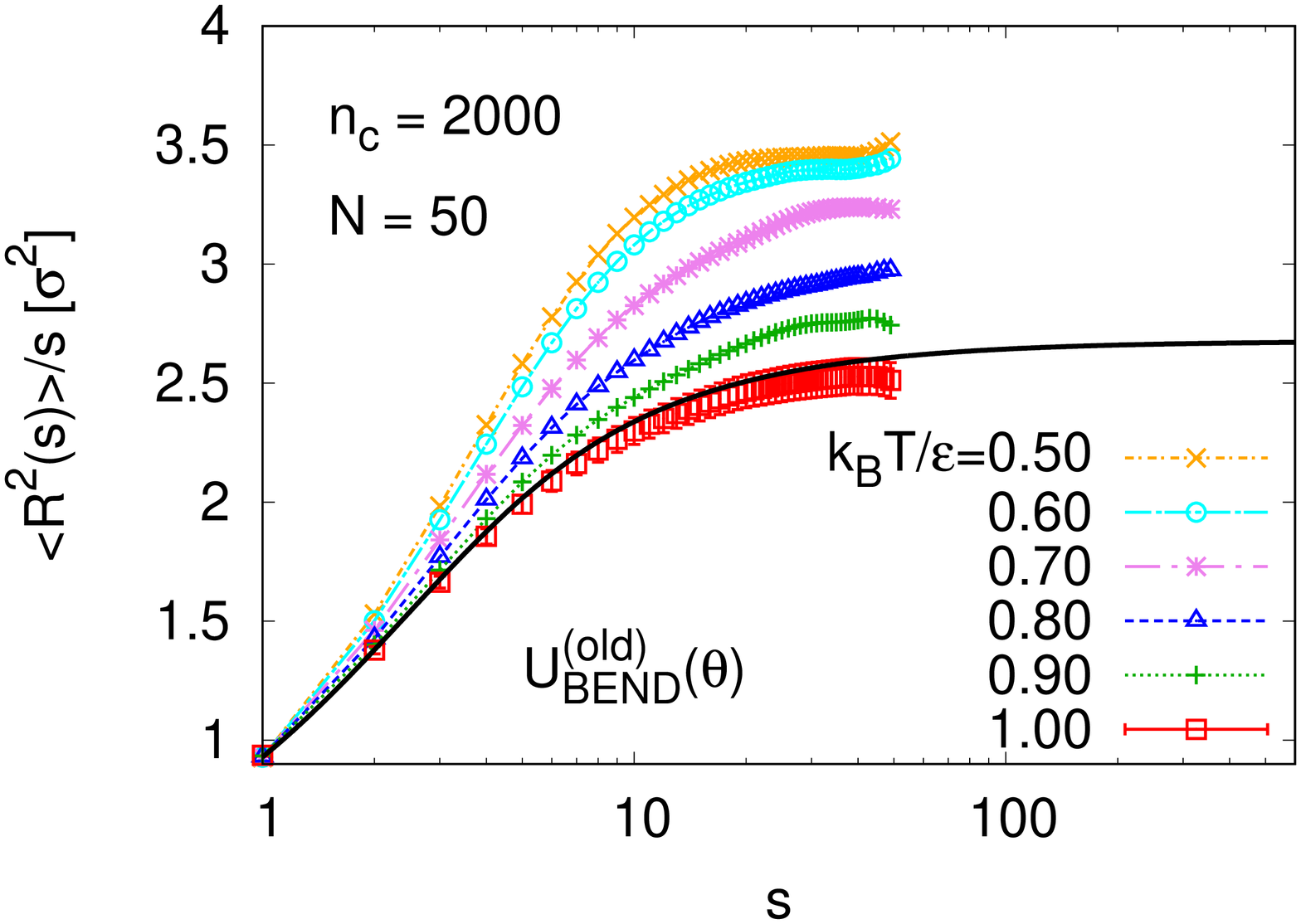} \hspace{1.2truecm}
(b)\includegraphics[width=0.30\textwidth,angle=270]{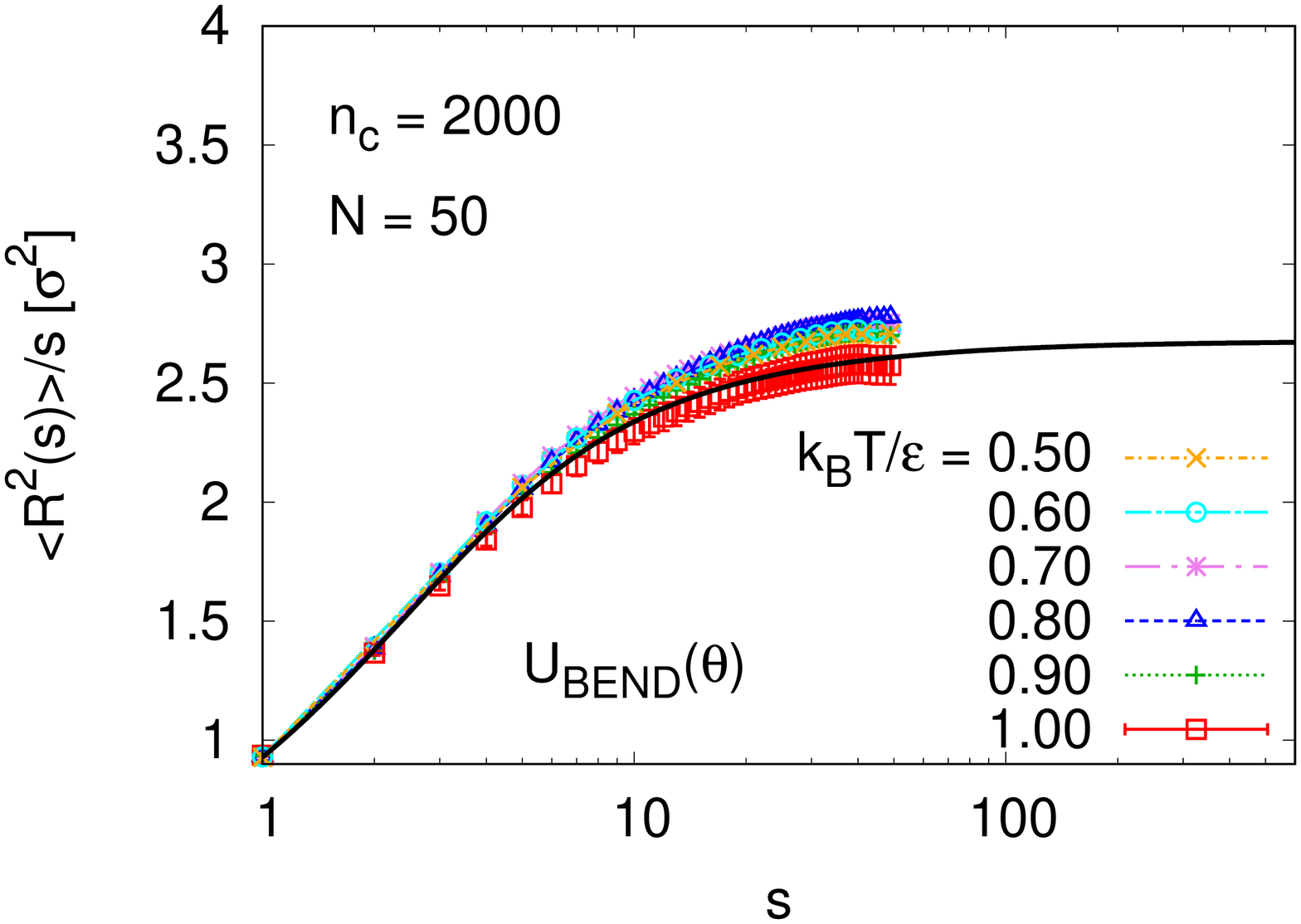}\\
(c)\includegraphics[width=0.30\textwidth,angle=270]{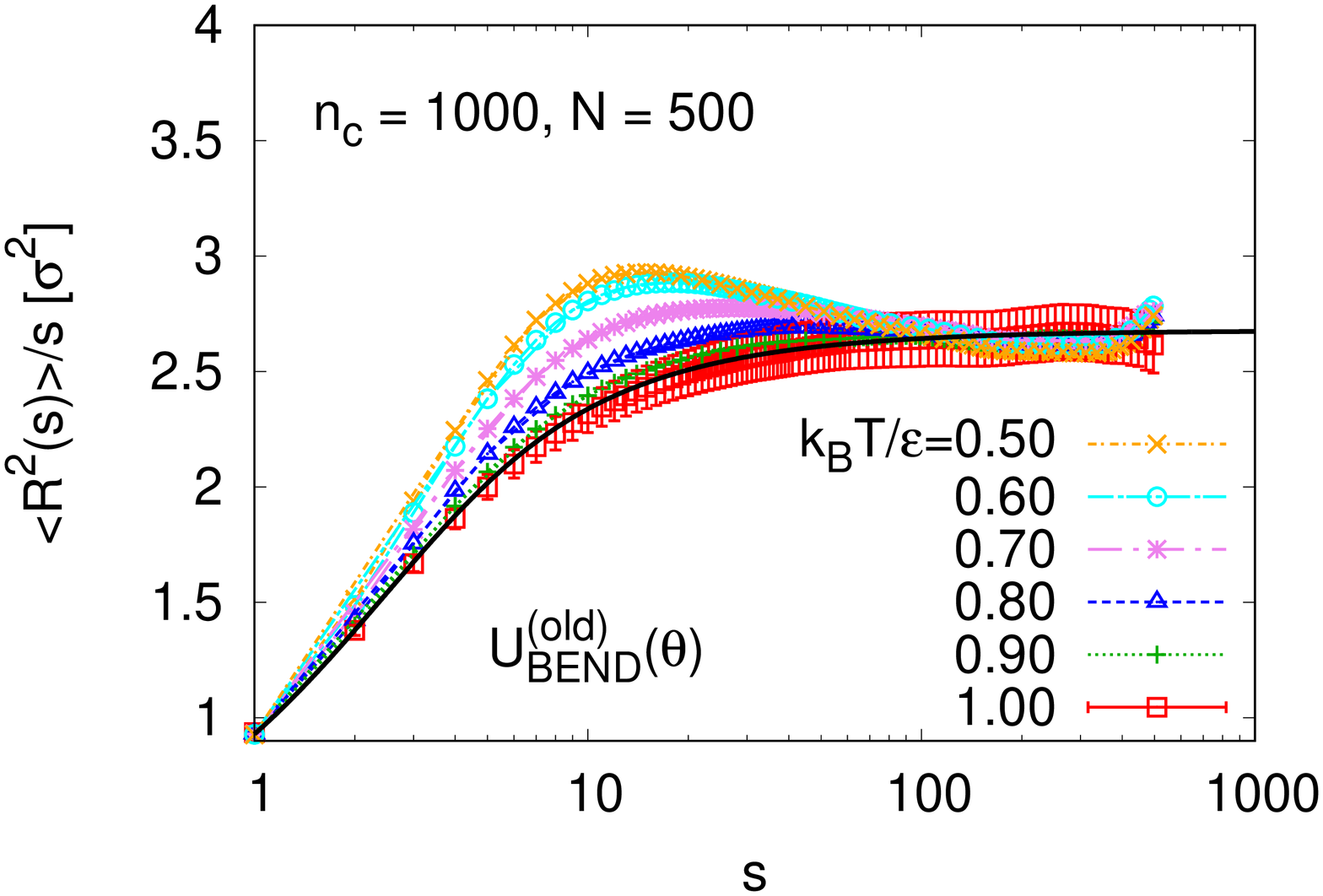} \hspace{1.2truecm}
(d)\includegraphics[width=0.30\textwidth,angle=270]{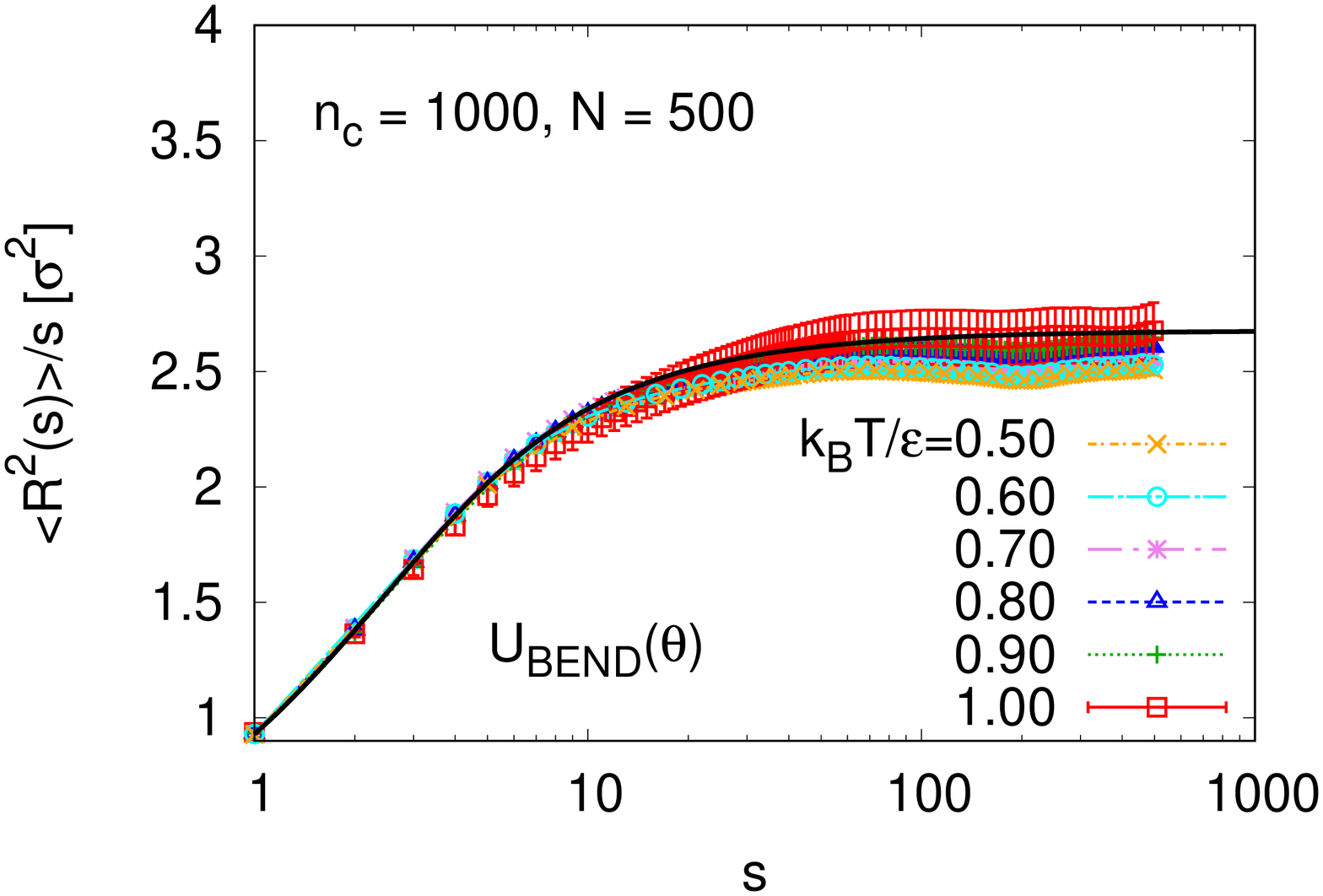}\\
\caption{Rescaled mean square internal distance, $\langle R^2(s) \rangle/s$,
plotted as a function of chemical distance $s$ for polymer melts described by
the standard BSM with the original and new bond bending potentials,
$U^{\rm (old)}_{\rm BEND}(\theta)$ (a)(c), and $U_{\rm BEND}(\theta)$ (b)(d), respectively,
at $P=0.0\epsilon/\sigma^3$.
The theoretical prediction for FRC with~\cite{Hsu2016} $\langle \cos \theta \rangle=0.4846$
estimated for fully equilibrated polymer melts of $n_c=1000$, $N=2000$
is also shown for comparison.}
\label{fig-R2s}
\end{center}
\end{figure*}

\begin{figure*}[t!]
\begin{center}
(a)\includegraphics[width=0.30\textwidth,angle=270]{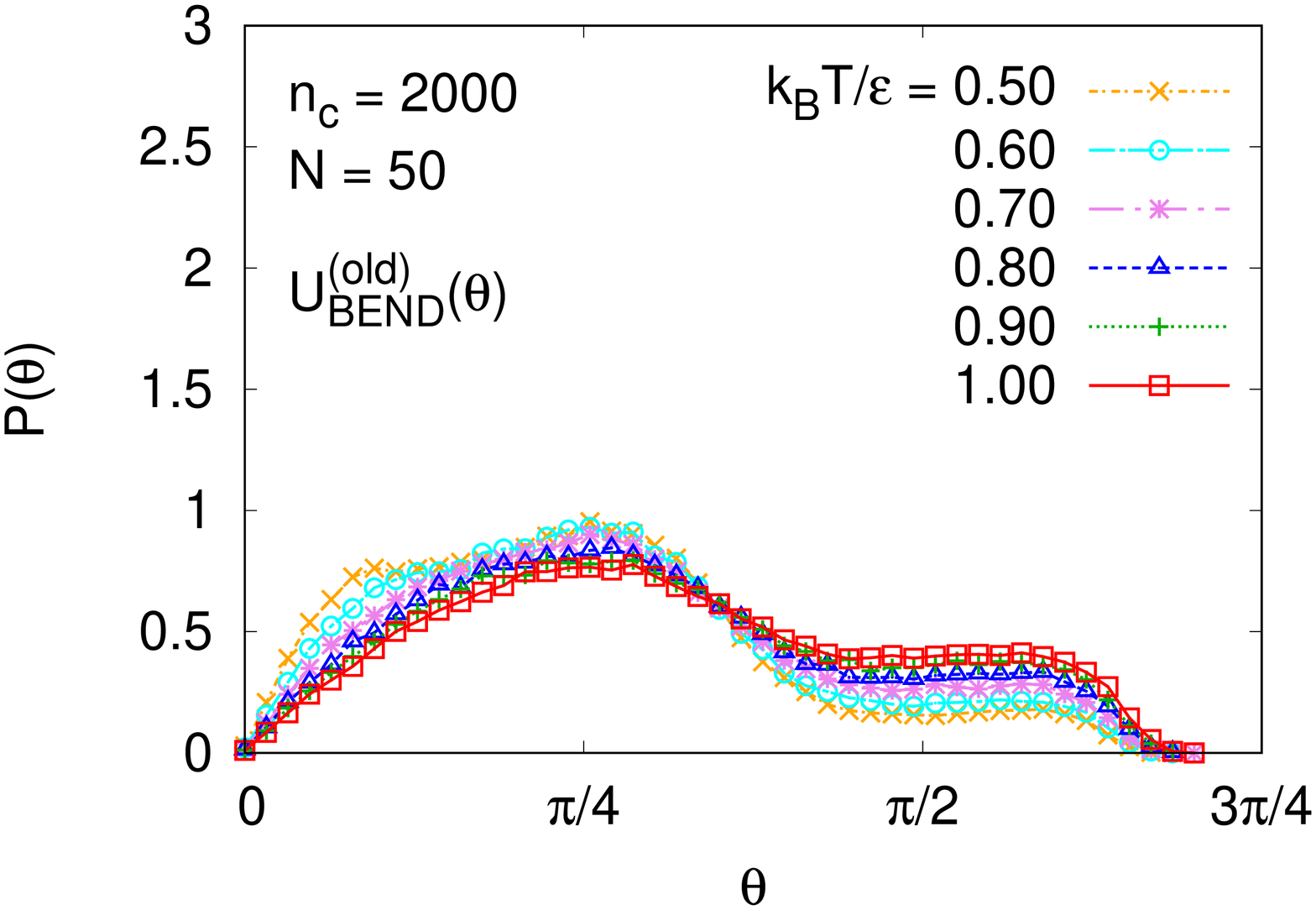} \hspace{1.2truecm}
(b)\includegraphics[width=0.30\textwidth,angle=270]{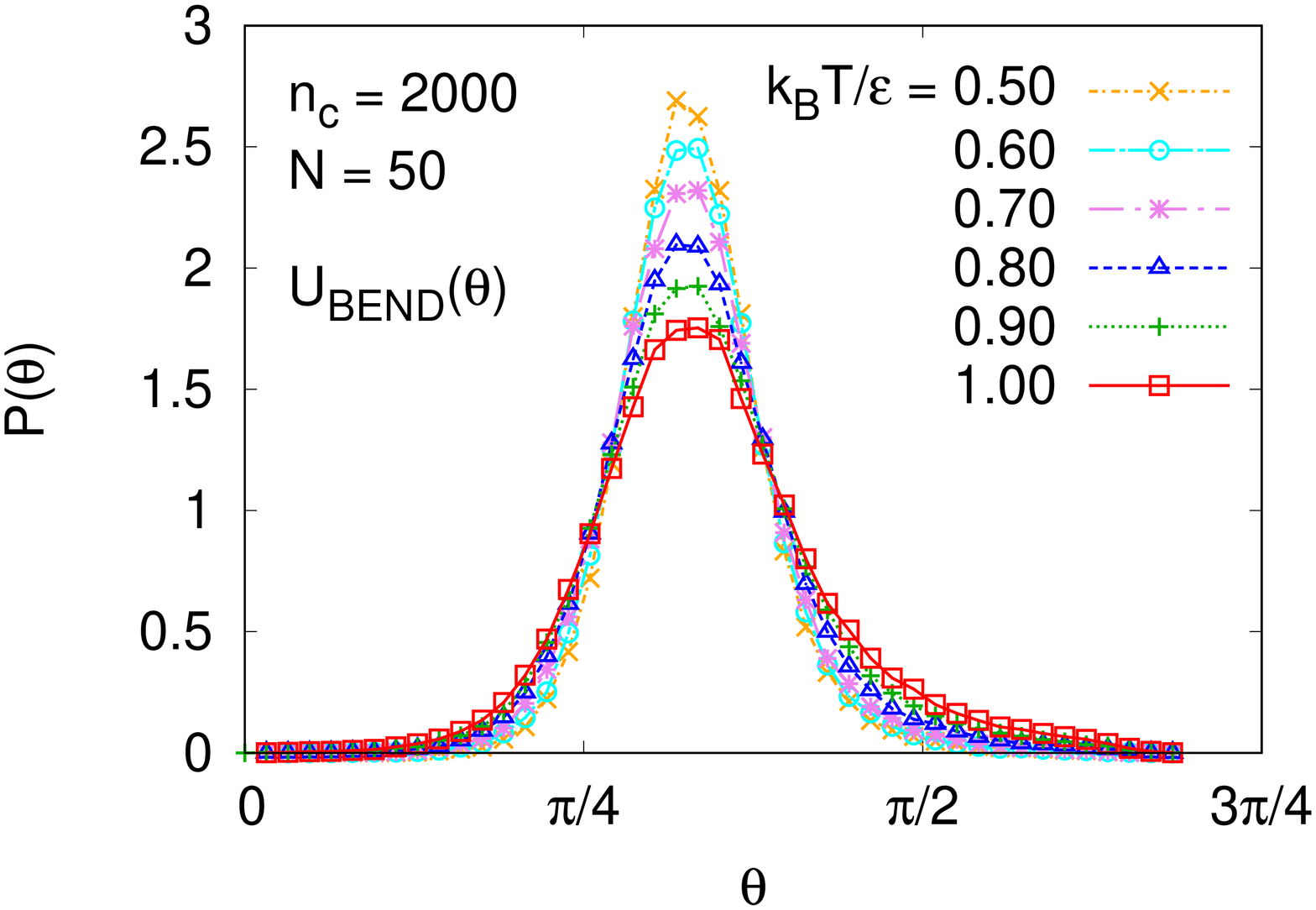}\\
\caption{Probability distribution of bond angle $\theta$ for polymer melts described by
the standard BSM with the original and new bond bending potential
$U^{\rm (old)}_{\rm BEND}(\theta)$ (a), and $U_{\rm BEND}(\theta)$
(b)), respectively.}
\label{fig-ptheta}
\end{center}
\end{figure*}

We now turn to the temperature dependency and compare melts of the new model to the standard 
semiflexible polymer model. For that we perform molecular dynamics (MD) simulations 
(Hoover Barostat with Langevin thermostat~\cite{Martyna1994, Quigley2004}
implemented in ESPResSo++~\cite{Espressopp}) at constant 
temperature $T$ by a stepwise cooling~\cite{Buchholz2002}, and constant 
pressure $P=0.0\epsilon/\sigma^3$ ($P=5.0\epsilon/\sigma^3$ for the old model), i.e. in the 
isothermal-isobaric ensemble (NPT), for two polymer melts of $n_c=2000$, $N=50$, 
and $n_c=1000$, $N=500$, respectively. The temperature is reduced in steps of 
$\Delta T=0.05\epsilon/k_B$ with a relaxation time  
between each step of $\Delta t=60000\tau$ resulting in
a cooling rate of $\Gamma=\Delta T/\Delta t=8.3 \times 10^{-7}\epsilon/(k_B\tau)$. 
$\Delta t$ corresponds to $ \approx 8.3 \tau_{R,N=50} \approx 0.083\tau_{R,N=500}$ 
($\tau_{R,N}$ being the Rouse time of the chains at $T=1.0\epsilon/k_B$
for the old model). 
Results of the mean square internal distances $\langle R^2(s) \rangle$ and the
bond angle probability distribution, $P(\theta)$, are shown in
Figures~\ref{fig-R2s}, \ref{fig-ptheta}.
First let us focus on the standard weakly semiflexible model. As temperature
decreases the chains stretch out as displayed in Figure~\ref{fig-R2s}a 
for $N=50$. While for $N=50$ the cooling rate is slow enough to allow for 
equilibration over a wide temperature range,  for longer chains ($N=500$, Figure~\ref{fig-R2s}c)
the system cannot equilibrate anymore even on short length scales ($s \leq 50$), leading to a
characteristic maximum in $\langle R^2(s) \rangle/s$. 
For long chain simulations, it will not be possible to avoid this artefact. Also the 
strong increase of $\langle R^2(s) \rangle$  of the standard 
semiflexible polymer model, is an artefact of the model when compared to experiments. 
This increase in chain stiffness is related to the shift of the probability
distribution $P(\theta)$ towards smaller angles as revealed in 
Figure~\ref{fig-ptheta}a and which directly connects to the shape of the 
standard bending potential~\cite{Schnell2011}.
In contrast, the new excluded volume and bending potential not only leads
to a conformational very close match with the old one at $T=1.0 \epsilon/k_B$, 
but  it also avoids a significant temperature shift.  
Figure~\ref{fig-ptheta}b demonstrates for $N=50$ that
$\langle R^2(s) \rangle /s$ becomes independent of $T$ within the error bars. 
As a consequence we also do not observe the maximum in 
$\langle R^2(s) \rangle$ for $N=500$ as a function of temperature (Figure~\ref{fig-R2s}d). 
These observations fit to the $T$ dependence of the distribution $P(\theta)$, 
Figure~\ref{fig-ptheta}b, which only becomes somewhat sharper but does not reveal 
any shift of the maximum.

\begin{figure*}[t!]
\begin{center}
(a)\includegraphics[width=0.30\textwidth,angle=270]{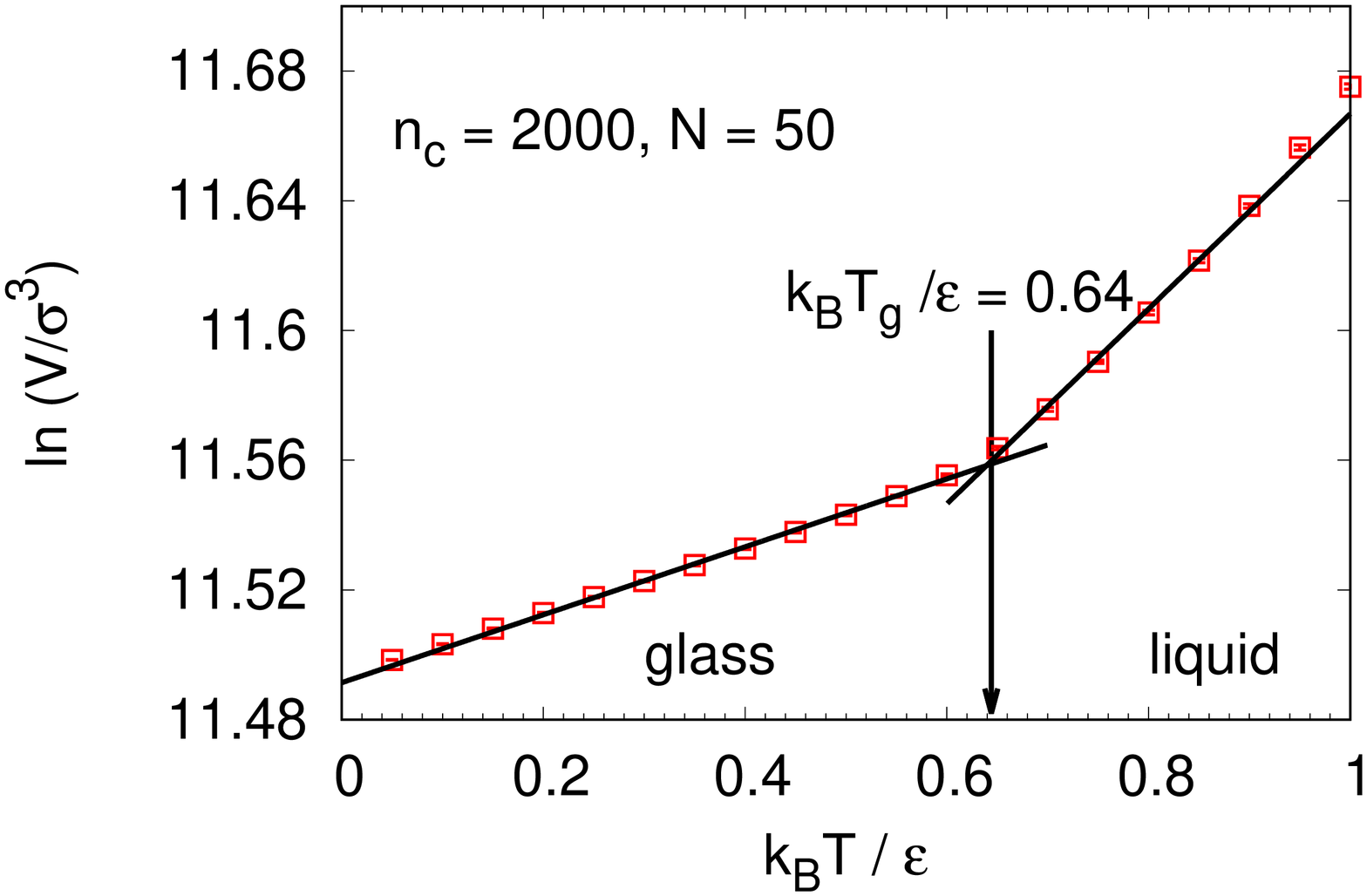} \hspace{1.2truecm}
(b)\includegraphics[width=0.30\textwidth,angle=270]{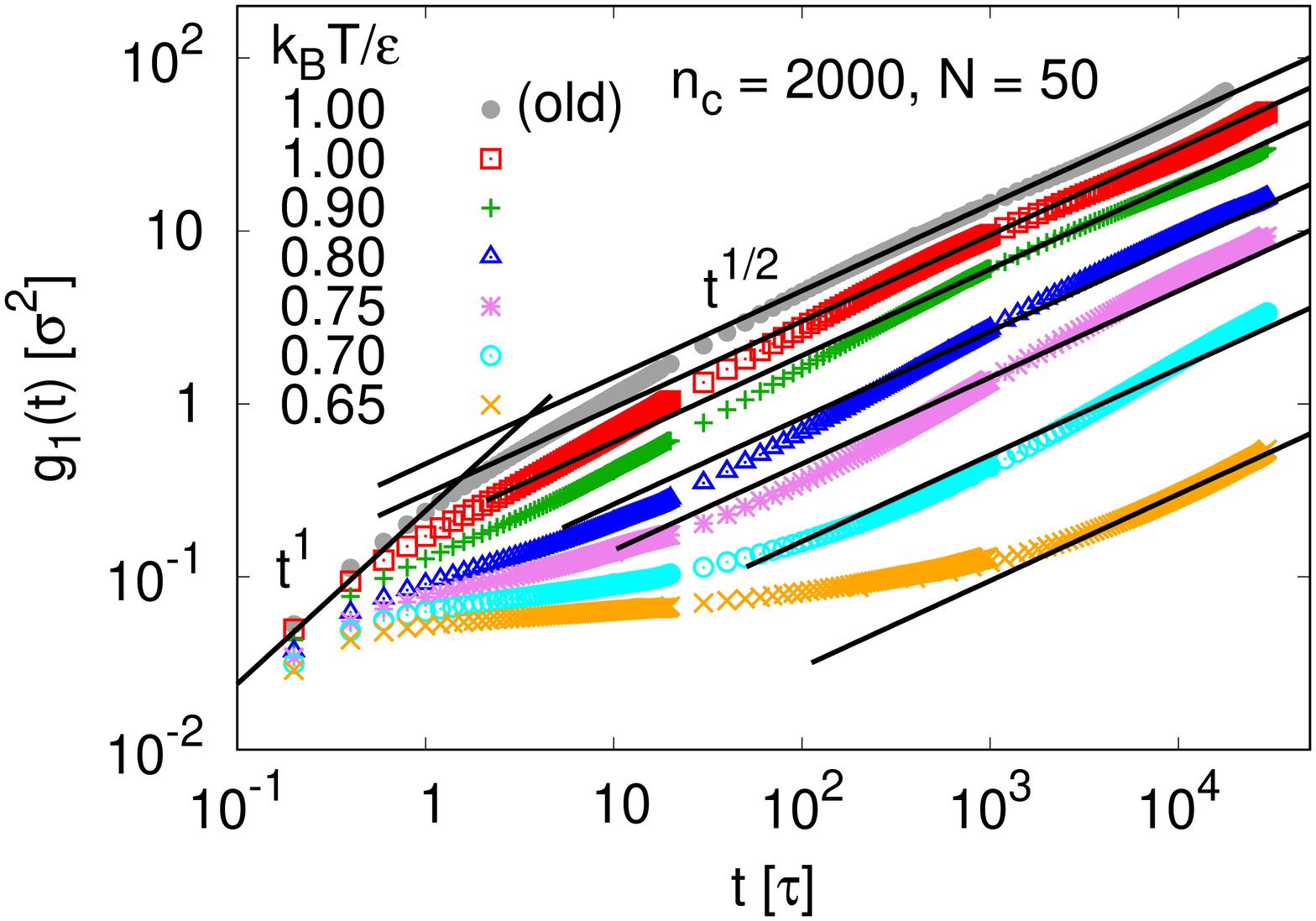}\\
(c)\includegraphics[width=0.30\textwidth,angle=270]{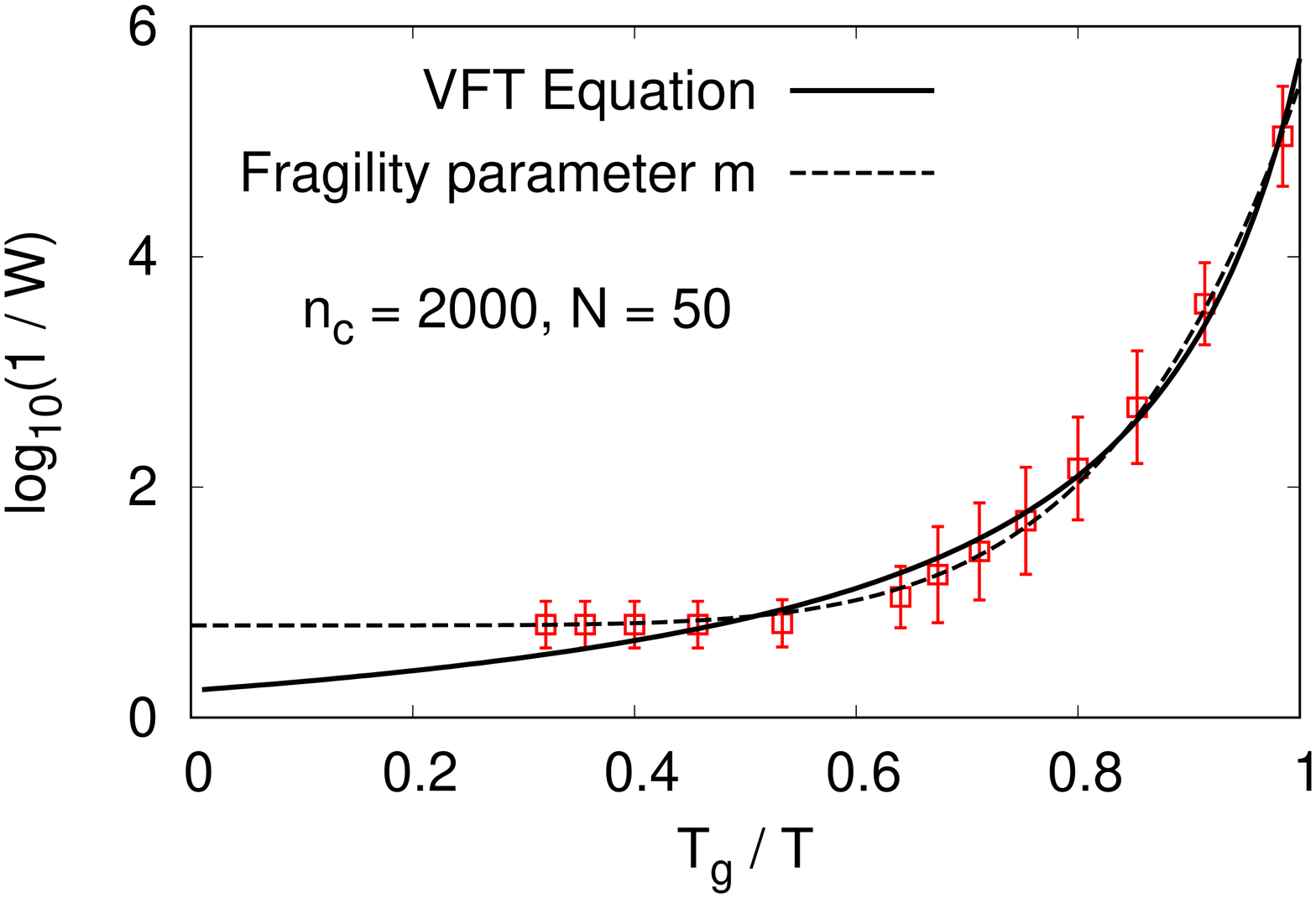}
\caption{(a) Logarithm of volume of the system, $\ln V/\sigma^3$, plotted versus temperature
$k_BT/\epsilon$. The two linear lines give the best fit of our data along the liquid branch
($a_{\rm liquid}=11.37$, $\alpha_{\rm liquid}=0.30k_B/\epsilon$) and the glass branch
($a_{\rm glass}=11.49$, $\alpha_{\rm glass}=0.10k_B/\epsilon$).
(b) Time evolution of mean square displacement of inner monomers, $g_1(t)$ at various chosen
temperatures $T$. The predicted scaling laws by Rouse model are shown by straight lines.
(c) Common logarithm of the inverse of the rate constant $W$ estimated from (b),
$\log_{10} (1/W)$, plotted versus $T_g/T$. Data for $k_BT/\epsilon>1.0$ are also included here.
The temperature dependence of the fragility parameter
$m(T)=4.7(T_g/T)^{6.0}+0.8$ is shown by a dashed curve,
and the VFT equation $\log_{10}(1/W)=A+B/(T-T_0)$ with
$A=0.24$, $B=0.45\epsilon/k_B$, and $T_0=0.56\epsilon/k_B$ is shown by a solid curve.}
\label{fig-Tg}
\end{center}
\end{figure*}

Finally we report some preliminary results for our new model in the glass transition region. 
As we are not interested here in details of the transition
itself, we focus on $N=50$ ($n_c=2000$) and one cooling rate ($\Gamma=8.3 \times 10^{-7} \epsilon/(k_B\tau))$, which, however,
allows for a full relaxation of the system up to the region very close
to $T_g$, the observed glass transition temperature.
$T_g$ can be determined from 
the change of density $\rho$ or volume $V$ as a function of temperature~\cite{Buchholz2002}. 
The intersection of linear extrapolation of $\ln V(T)$
between the liquid branch 
($\ln V_{\rm liquid}=a_{\rm liquid}+\alpha_{\rm liquid}T$) and 
glass branch ($\ln V_{\rm glass}=a_{\rm glass}+\alpha_{\rm glass}T$) 
gives a good estimate of $T_g$.
Here $\alpha_{\rm liquid}$ and $\alpha_{\rm glass}$ are thermal expansion
coefficients for polymer melts in the liquid state and the glass state, 
respectively. 
Results of $\ln V$ plotted versus $T$ are shown in 
Figure~\ref{fig-Tg}a. The glass transition occurs around $T_g=0.64\epsilon/k_B$.
To investigate the mobility of chains at $T>T_g$, we perform additional NVT MD simulations
with a weak coupling Langevin thermostat for polymer melts 
at $k_BT/\epsilon=1.0$, $0.95$, $0.90$, $0.85$, $0.80$,
$0.75$, $0.70$, and $0.65$. The initial configuration and volume of the polymer
melt at each temperature $T$ are taken from the last configuration of the 
NPT run in the cooling process. According to the Rouse model~\cite{Rouse1953}, 
the mean square displacement (MSD) of monomers, $g_1(t)$, is expressed in terms 
of the Rouse rate $W=12k_BT/(\pi \zeta \sigma^2)$ as $g_1(t)=\sigma^2(Wt)^{1/2}$.
Here $\zeta (\propto D^{-1} \propto \eta)$ being the monomeric friction 
coefficient is related to the self-diffusion coefficient $D=k_BT/(N\zeta)$
and the viscosity $\eta$ using the Stokes-Einstein relation.
Results of $g_1(t)$ taking from the
average MSD of inner $12$ monomers are shown in Figure~\ref{fig-Tg}b.
We also include the data at $T=1.0\epsilon/k_B$ for the old model for comparison.
The Rouse rate $W$ depending on the temperature is determined by the best
fit of a straight line with slope $1/2$ going through our data on log-log scales.
At $T=1.0\epsilon/k_B$, the Rouse rate for the old model ($W=0.20\tau^{-1}$) is faster than
the new model ($W=0.09\tau^{-1}$).
From the well-known Vogel-Fulcher-Tammann (VFT) equation~\cite{Vogel1921,Fulcher1925,Tammann1926},
$\log_{10} \eta = A+ \frac{B}{T-T_0}$, where $A$, $B$, and $T_0$ are constants and $T$ 
is the absolute temperature, Angell~\cite{Angell1988,Angell1991} has
proposed that the fragility parameter $m$, defined by~\cite{Nascimento2007}:
$m=d(\log_{10} \eta)/d(T_g/T) \mid_{T=T_g}$. Thus, plotting $\log_{10}(1/W)$ 
versus $T_g/T$ in Figure~\ref{fig-Tg}c, we obtain the characteristic behavior of a polymer 
approaching the glass transition.

In summary, based on the standard BSM, we have introduced a new non-bonded short range
attractive potential $U_{\rm ATT}(r)$ and bond bending potential $U_{\rm BEND}(\theta)$
for studying polymer melts subject to cooling. The functional form of these two new 
interaction potentials also is directly applicable to other standard BSM models with
different stiffness~\cite{Svaneborg2018a} just by adjusting the coefficients.
By keeping $\alpha=0.5145\epsilon$, which results in a density of $0.85\sigma^{-3}$ for 
all longest ($N=2000$) systems within the error bars, we get 
$a_\theta=4.5\epsilon$ for $0\le k_\theta/\epsilon \le 2.0$, and 
$b_\theta=1.32$, $1.40$, $1.50$, and $1.70$ for $k_\theta/\epsilon=0.5$, $1.0$, $1.5$, and $2.0$, respectively.
The new coarse-grained model captures 
the major features of glass-forming polymers, and preserves the Kuhn length as well as 
internal distances and can also be used to study systems with free surfaces. By construction it can directly take advantage of available simulation data 
of standard BSM models at $T=1.0 \epsilon/k_B$ and can be applied to available large 
deformed polymer melts~\cite{Hsu2018a,Hsu2018b} and for understanding the viscoelastic behavior of these polymeric systems. 

Acknowledgement: 
We are grateful to B. D\"unweg for a critical reading of the manuscript.
This work has been supported by European Research Council under the European
Union's Seventh Framework Programme (FP7/2007-2013)/ERC Grant Agreement
No.~340906-MOLPROCOMP.
We also gratefully acknowledge the computing time granted by the John von
Neumann Institute for Computing (NIC) and provided on the supercomputer JUROPA
at J\"ulich Supercomputing Centre (JSC),
and the Max Planck Computing and Data Facility (MPCDF).

%\bibliography{Ref_2018.bib}

\begin{thebibliography}{43}%
\makeatletter
\providecommand \@ifxundefined [1]{%
 \@ifx{#1\undefined}
}%
\providecommand \@ifnum [1]{%
 \ifnum #1\expandafter \@firstoftwo
 \else \expandafter \@secondoftwo
 \fi
}%
\providecommand \@ifx [1]{%
 \ifx #1\expandafter \@firstoftwo
 \else \expandafter \@secondoftwo
 \fi
}%
\providecommand \natexlab [1]{#1}%
\providecommand \enquote  [1]{``#1''}%
\providecommand \bibnamefont  [1]{#1}%
\providecommand \bibfnamefont [1]{#1}%
\providecommand \citenamefont [1]{#1}%
\providecommand \href@noop [0]{\@secondoftwo}%
\providecommand \href [0]{\begingroup \@sanitize@url \@href}%
\providecommand \@href[1]{\@@startlink{#1}\@@href}%
\providecommand \@@href[1]{\endgroup#1\@@endlink}%
\providecommand \@sanitize@url [0]{\catcode `\\12\catcode `\$12\catcode
  `\&12\catcode `\#12\catcode `\^12\catcode `\_12\catcode `\%12\relax}%
\providecommand \@@startlink[1]{}%
\providecommand \@@endlink[0]{}%
\providecommand \url  [0]{\begingroup\@sanitize@url \@url }%
\providecommand \@url [1]{\endgroup\@href {#1}{\urlprefix }}%
\providecommand \urlprefix  [0]{URL }%
\providecommand \Eprint [0]{\href }%
\providecommand \doibase [0]{http://dx.doi.org/}%
\providecommand \selectlanguage [0]{\@gobble}%
\providecommand \bibinfo  [0]{\@secondoftwo}%
\providecommand \bibfield  [0]{\@secondoftwo}%
\providecommand \translation [1]{[#1]}%
\providecommand \BibitemOpen [0]{}%
\providecommand \bibitemStop [0]{}%
\providecommand \bibitemNoStop [0]{.\EOS\space}%
\providecommand \EOS [0]{\spacefactor3000\relax}%
\providecommand \BibitemShut  [1]{\csname bibitem#1\endcsname}%
\let\auto@bib@innerbib\@empty
%</preamble>
\bibitem [{\citenamefont {Anantrao}\ \emph {et~al.}(2017)\citenamefont
  {Anantrao}, \citenamefont {Motichand},\ and\ \citenamefont
  {Narhari}}]{Hrushikesh2017}%
  \BibitemOpen
  \bibfield  {author} {\bibinfo {author} {\bibfnamefont {J.~H.}\ \bibnamefont
  {Anantrao}}, \bibinfo {author} {\bibfnamefont {J.~C.}\ \bibnamefont
  {Motichand}}, \ and\ \bibinfo {author} {\bibfnamefont {B.~S.}\ \bibnamefont
  {Narhari}},\ }\href@noop {} {\bibfield  {journal} {\bibinfo  {journal} {Int.
  J. Adv. Res.}\ }\textbf {\bibinfo {volume} {5}},\ \bibinfo {pages} {671}
  (\bibinfo {year} {2017})}\BibitemShut {NoStop}%
\bibitem [{\citenamefont {Boyd}\ \emph {et~al.}(1994)\citenamefont {Boyd},
  \citenamefont {Gee}, \citenamefont {Han},\ and\ \citenamefont
  {Jin}}]{Boyd1994}%
  \BibitemOpen
  \bibfield  {author} {\bibinfo {author} {\bibfnamefont {R.~H.}\ \bibnamefont
  {Boyd}}, \bibinfo {author} {\bibfnamefont {R.~H.}\ \bibnamefont {Gee}},
  \bibinfo {author} {\bibfnamefont {J.}~\bibnamefont {Han}}, \ and\ \bibinfo
  {author} {\bibfnamefont {Y.}~\bibnamefont {Jin}},\ }\href@noop {} {\bibfield
  {journal} {\bibinfo  {journal} {J. Chem. Phys.}\ }\textbf {\bibinfo {volume}
  {101}},\ \bibinfo {pages} {788} (\bibinfo {year} {1994})}\BibitemShut
  {NoStop}%
\bibitem [{\citenamefont {Angell}(1995)}]{Angell1995}%
  \BibitemOpen
  \bibfield  {author} {\bibinfo {author} {\bibfnamefont {C.~A.}\ \bibnamefont
  {Angell}},\ }\href@noop {} {\bibfield  {journal} {\bibinfo  {journal}
  {Science}\ }\textbf {\bibinfo {volume} {267}},\ \bibinfo {pages} {1924}
  (\bibinfo {year} {1995})}\BibitemShut {NoStop}%
\bibitem [{\citenamefont {Paluch}\ \emph {et~al.}(2001)\citenamefont {Paluch},
  \citenamefont {Gapinski}, \citenamefont {Patkowski},\ and\ \citenamefont
  {Fischer}}]{Paluch2001}%
  \BibitemOpen
  \bibfield  {author} {\bibinfo {author} {\bibfnamefont {M.}~\bibnamefont
  {Paluch}}, \bibinfo {author} {\bibfnamefont {J.}~\bibnamefont {Gapinski}},
  \bibinfo {author} {\bibfnamefont {A.}~\bibnamefont {Patkowski}}, \ and\
  \bibinfo {author} {\bibfnamefont {E.~W.}\ \bibnamefont {Fischer}},\
  }\href@noop {} {\bibfield  {journal} {\bibinfo  {journal} {J. Chem. Phys.}\
  }\textbf {\bibinfo {volume} {114}},\ \bibinfo {pages} {8048} (\bibinfo {year}
  {2001})}\BibitemShut {NoStop}%
\bibitem [{\citenamefont {Berthier}\ and\ \citenamefont
  {Biroli}(2011)}]{Berthier2011}%
  \BibitemOpen
  \bibfield  {author} {\bibinfo {author} {\bibfnamefont {L.}~\bibnamefont
  {Berthier}}\ and\ \bibinfo {author} {\bibfnamefont {G.}~\bibnamefont
  {Biroli}},\ }\href@noop {} {\bibfield  {journal} {\bibinfo  {journal} {Rev.
  Mod. Phys.}\ }\textbf {\bibinfo {volume} {83}},\ \bibinfo {pages} {587}
  (\bibinfo {year} {2011})}\BibitemShut {NoStop}%
\bibitem [{\citenamefont {Mathot}(1994)}]{Mathot1994}%
  \BibitemOpen
  \bibfield  {author} {\bibinfo {author} {\bibfnamefont {V.~B.~F.}\
  \bibnamefont {Mathot}},\ }\href@noop {} {\emph {\bibinfo {title} {Calorimetry
  and thermal analysis of polymers}}}\ (\bibinfo  {publisher} {Hanser,
  Munich},\ \bibinfo {year} {1994})\BibitemShut {NoStop}%
\bibitem [{\citenamefont {Bird}\ \emph {et~al.}(1987)\citenamefont {Bird},
  \citenamefont {Curtis},\ and\ \citenamefont {Armstrong}}]{Bird1987}%
  \BibitemOpen
  \bibfield  {author} {\bibinfo {author} {\bibfnamefont {R.}~\bibnamefont
  {Bird}}, \bibinfo {author} {\bibfnamefont {C.}~\bibnamefont {Curtis}}, \ and\
  \bibinfo {author} {\bibfnamefont {R.}~\bibnamefont {Armstrong}},\ }\href@noop
  {} {\emph {\bibinfo {title} {Dynamics of Polymer Fluids, 2nd ed.}}}\
  (\bibinfo  {publisher} {Wiley, New York},\ \bibinfo {year}
  {1987})\BibitemShut {NoStop}%
\bibitem [{\citenamefont {Angell}(1988)}]{Angell1988}%
  \BibitemOpen
  \bibfield  {author} {\bibinfo {author} {\bibfnamefont {C.~A.}\ \bibnamefont
  {Angell}},\ }\href@noop {} {\bibfield  {journal} {\bibinfo  {journal} {J.
  Phys. Chem. Solids}\ }\textbf {\bibinfo {volume} {49}},\ \bibinfo {pages}
  {863} (\bibinfo {year} {1988})}\BibitemShut {NoStop}%
\bibitem [{\citenamefont {Angell}(1991)}]{Angell1991}%
  \BibitemOpen
  \bibfield  {author} {\bibinfo {author} {\bibfnamefont {C.~A.}\ \bibnamefont
  {Angell}},\ }\href {\doibase 10.1016/0022-3093(91)90266-9} {\bibfield
  {journal} {\bibinfo  {journal} {J. Non-Cryst. Solids}\ }\textbf {\bibinfo
  {volume} {131-133, Part 1}},\ \bibinfo {pages} {13} (\bibinfo {year}
  {1991})}\BibitemShut {NoStop}%
\bibitem [{\citenamefont {Binder}\ and\ \citenamefont
  {Kob}(2005)}]{Binder2005}%
  \BibitemOpen
  \bibfield  {author} {\bibinfo {author} {\bibfnamefont {K.}~\bibnamefont
  {Binder}}\ and\ \bibinfo {author} {\bibfnamefont {W.}~\bibnamefont {Kob}},\
  }\href@noop {} {\emph {\bibinfo {title} {Glassy Materials and Disordered
  Solids}}}\ (\bibinfo  {publisher} {World Scientific, Singapore},\ \bibinfo
  {year} {2005})\BibitemShut {NoStop}%
\bibitem [{\citenamefont {Barrat}\ \emph {et~al.}(2010)\citenamefont {Barrat},
  \citenamefont {Baschnagel},\ and\ \citenamefont {Lyulin}}]{Barrat2010}%
  \BibitemOpen
  \bibfield  {author} {\bibinfo {author} {\bibfnamefont {J.-L.}\ \bibnamefont
  {Barrat}}, \bibinfo {author} {\bibfnamefont {J.}~\bibnamefont {Baschnagel}},
  \ and\ \bibinfo {author} {\bibfnamefont {A.}~\bibnamefont {Lyulin}},\
  }\href@noop {} {\bibfield  {journal} {\bibinfo  {journal} {Soft Matter}\
  }\textbf {\bibinfo {volume} {6}},\ \bibinfo {pages} {3430} (\bibinfo {year}
  {2010})}\BibitemShut {NoStop}%
\bibitem [{\citenamefont {Stillinger}\ and\ \citenamefont
  {Debenedetti}(2013)}]{Stillinger2013}%
  \BibitemOpen
  \bibfield  {author} {\bibinfo {author} {\bibfnamefont {F.~H.}\ \bibnamefont
  {Stillinger}}\ and\ \bibinfo {author} {\bibfnamefont {P.~G.}\ \bibnamefont
  {Debenedetti}},\ }\href@noop {} {\bibfield  {journal} {\bibinfo  {journal}
  {Annu. Rev. Condens. Matter Phys.}\ }\textbf {\bibinfo {volume} {4}},\
  \bibinfo {pages} {263} (\bibinfo {year} {2013})}\BibitemShut {NoStop}%
\bibitem [{\citenamefont {Ediger}\ and\ \citenamefont
  {Forrest}(2014)}]{Ediger2014}%
  \BibitemOpen
  \bibfield  {author} {\bibinfo {author} {\bibfnamefont {M.~D.}\ \bibnamefont
  {Ediger}}\ and\ \bibinfo {author} {\bibfnamefont {J.~A.}\ \bibnamefont
  {Forrest}},\ }\href@noop {} {\bibfield  {journal} {\bibinfo  {journal}
  {Macromolecules}\ }\textbf {\bibinfo {volume} {47}},\ \bibinfo {pages} {471}
  (\bibinfo {year} {2014})}\BibitemShut {NoStop}%
\bibitem [{\citenamefont {Chaimovich}\ \emph {et~al.}(2015)\citenamefont
  {Chaimovich}, \citenamefont {Peter},\ and\ \citenamefont
  {Kremer}}]{Chaimovich2015}%
  \BibitemOpen
  \bibfield  {author} {\bibinfo {author} {\bibfnamefont {A.}~\bibnamefont
  {Chaimovich}}, \bibinfo {author} {\bibfnamefont {C.}~\bibnamefont {Peter}}, \
  and\ \bibinfo {author} {\bibfnamefont {K.}~\bibnamefont {Kremer}},\
  }\href@noop {} {\bibfield  {journal} {\bibinfo  {journal} {J. Chem. Phys.}\
  }\textbf {\bibinfo {volume} {143}},\ \bibinfo {pages} {243107} (\bibinfo
  {year} {2015})}\BibitemShut {NoStop}%
\bibitem [{\citenamefont {Kremer}\ and\ \citenamefont
  {Grest}(1990)}]{Kremer1990}%
  \BibitemOpen
  \bibfield  {author} {\bibinfo {author} {\bibfnamefont {K.}~\bibnamefont
  {Kremer}}\ and\ \bibinfo {author} {\bibfnamefont {G.~S.}\ \bibnamefont
  {Grest}},\ }\href@noop {} {\bibfield  {journal} {\bibinfo  {journal} {J.
  Chem. Phys.}\ }\textbf {\bibinfo {volume} {92}},\ \bibinfo {pages} {5057}
  (\bibinfo {year} {1990})}\BibitemShut {NoStop}%
\bibitem [{\citenamefont {D\"unweg}\ \emph {et~al.}(1997)\citenamefont
  {D\"unweg}, \citenamefont {Grest},\ and\ \citenamefont
  {Kremer}}]{Duenweg1997}%
  \BibitemOpen
  \bibfield  {author} {\bibinfo {author} {\bibfnamefont {B.}~\bibnamefont
  {D\"unweg}}, \bibinfo {author} {\bibfnamefont {G.~S.}\ \bibnamefont {Grest}},
  \ and\ \bibinfo {author} {\bibfnamefont {K.}~\bibnamefont {Kremer}},\
  }\href@noop {} {\emph {\bibinfo {title} {Conf. Proc. of the IMA Workshop
  (Minneapolis MN, May 1996)}}}\ (\bibinfo  {publisher} {Berlin:Springer},\
  \bibinfo {year} {1997})\BibitemShut {NoStop}%
\bibitem [{\citenamefont {Kopf}\ \emph {et~al.}(1997)\citenamefont {Kopf},
  \citenamefont {D\"unweg},\ and\ \citenamefont {Paul}}]{Kopf1997}%
  \BibitemOpen
  \bibfield  {author} {\bibinfo {author} {\bibfnamefont {A.}~\bibnamefont
  {Kopf}}, \bibinfo {author} {\bibfnamefont {B.}~\bibnamefont {D\"unweg}}, \
  and\ \bibinfo {author} {\bibfnamefont {W.}~\bibnamefont {Paul}},\ }\href@noop
  {} {\bibfield  {journal} {\bibinfo  {journal} {J. Chem. Phys.}\ }\textbf
  {\bibinfo {volume} {107}},\ \bibinfo {pages} {6945} (\bibinfo {year}
  {1997})}\BibitemShut {NoStop}%
\bibitem [{\citenamefont {Bennemann}\ \emph {et~al.}(1998)\citenamefont
  {Bennemann}, \citenamefont {Paul}, \citenamefont {Binder},\ and\
  \citenamefont {D\"unweg}}]{Bennemann1998}%
  \BibitemOpen
  \bibfield  {author} {\bibinfo {author} {\bibfnamefont {C.}~\bibnamefont
  {Bennemann}}, \bibinfo {author} {\bibfnamefont {W.}~\bibnamefont {Paul}},
  \bibinfo {author} {\bibfnamefont {K.}~\bibnamefont {Binder}}, \ and\ \bibinfo
  {author} {\bibfnamefont {B.}~\bibnamefont {D\"unweg}},\ }\href@noop {}
  {\bibfield  {journal} {\bibinfo  {journal} {Phys. Rev. E}\ }\textbf {\bibinfo
  {volume} {57}},\ \bibinfo {pages} {843} (\bibinfo {year} {1998})}\BibitemShut
  {NoStop}%
\bibitem [{\citenamefont {Binder}(1999)}]{Binder1999}%
  \BibitemOpen
  \bibfield  {author} {\bibinfo {author} {\bibfnamefont {K.}~\bibnamefont
  {Binder}},\ }\href@noop {} {\bibfield  {journal} {\bibinfo  {journal} {Comp.
  Phys. Commun.}\ }\textbf {\bibinfo {volume} {121-122}},\ \bibinfo {pages}
  {168} (\bibinfo {year} {1999})}\BibitemShut {NoStop}%
\bibitem [{\citenamefont {Buchholz}\ \emph {et~al.}(2002)\citenamefont
  {Buchholz}, \citenamefont {Paul}, \citenamefont {Varnik},\ and\ \citenamefont
  {Binder}}]{Buchholz2002}%
  \BibitemOpen
  \bibfield  {author} {\bibinfo {author} {\bibfnamefont {J.}~\bibnamefont
  {Buchholz}}, \bibinfo {author} {\bibfnamefont {W.}~\bibnamefont {Paul}},
  \bibinfo {author} {\bibfnamefont {F.}~\bibnamefont {Varnik}}, \ and\ \bibinfo
  {author} {\bibfnamefont {K.}~\bibnamefont {Binder}},\ }\href@noop {}
  {\bibfield  {journal} {\bibinfo  {journal} {J. Chem. Phys.}\ }\textbf
  {\bibinfo {volume} {117}},\ \bibinfo {pages} {7364} (\bibinfo {year}
  {2002})}\BibitemShut {NoStop}%
\bibitem [{\citenamefont {Binder}\ \emph {et~al.}(2003)\citenamefont {Binder},
  \citenamefont {Baschnagel},\ and\ \citenamefont {Paul}}]{Binder2003}%
  \BibitemOpen
  \bibfield  {author} {\bibinfo {author} {\bibfnamefont {K.}~\bibnamefont
  {Binder}}, \bibinfo {author} {\bibfnamefont {J.}~\bibnamefont {Baschnagel}},
  \ and\ \bibinfo {author} {\bibfnamefont {W.}~\bibnamefont {Paul}},\
  }\href@noop {} {\bibfield  {journal} {\bibinfo  {journal} {Prog. Polym.
  Sci.}\ }\textbf {\bibinfo {volume} {28}},\ \bibinfo {pages} {115} (\bibinfo
  {year} {2003})}\BibitemShut {NoStop}%
\bibitem [{\citenamefont {Schnell}\ \emph {et~al.}(2011)\citenamefont
  {Schnell}, \citenamefont {Meyer}, \citenamefont {Fond}, \citenamefont
  {Wittmer},\ and\ \citenamefont {Baschnagel}}]{Schnell2011}%
  \BibitemOpen
  \bibfield  {author} {\bibinfo {author} {\bibfnamefont {B.}~\bibnamefont
  {Schnell}}, \bibinfo {author} {\bibfnamefont {H.}~\bibnamefont {Meyer}},
  \bibinfo {author} {\bibfnamefont {C.}~\bibnamefont {Fond}}, \bibinfo {author}
  {\bibfnamefont {J.~P.}\ \bibnamefont {Wittmer}}, \ and\ \bibinfo {author}
  {\bibfnamefont {J.}~\bibnamefont {Baschnagel}},\ }\href@noop {} {\bibfield
  {journal} {\bibinfo  {journal} {Eur. Phys. J. E}\ }\textbf {\bibinfo {volume}
  {34}},\ \bibinfo {pages} {97} (\bibinfo {year} {2011})}\BibitemShut {NoStop}%
\bibitem [{\citenamefont {Frey}\ \emph {et~al.}(2015)\citenamefont {Frey},
  \citenamefont {Weysser}, \citenamefont {Meyer}, \citenamefont {Farago},
  \citenamefont {Fuchs},\ and\ \citenamefont {Baschnagel}}]{Frey2015}%
  \BibitemOpen
  \bibfield  {author} {\bibinfo {author} {\bibfnamefont {S.}~\bibnamefont
  {Frey}}, \bibinfo {author} {\bibfnamefont {F.}~\bibnamefont {Weysser}},
  \bibinfo {author} {\bibfnamefont {H.}~\bibnamefont {Meyer}}, \bibinfo
  {author} {\bibfnamefont {J.}~\bibnamefont {Farago}}, \bibinfo {author}
  {\bibfnamefont {M.}~\bibnamefont {Fuchs}}, \ and\ \bibinfo {author}
  {\bibfnamefont {J.}~\bibnamefont {Baschnagel}},\ }\href@noop {} {\bibfield
  {journal} {\bibinfo  {journal} {Eur. Phys. J. E.}\ }\textbf {\bibinfo
  {volume} {38}},\ \bibinfo {pages} {11} (\bibinfo {year} {2015})}\BibitemShut
  {NoStop}%
\bibitem [{\citenamefont {Everaers}\ \emph {et~al.}(2004)\citenamefont
  {Everaers}, \citenamefont {Sukumaran}, \citenamefont {Grest}, \citenamefont
  {Svaneborg}, \citenamefont {Sivasubramanian},\ and\ \citenamefont
  {Kremer}}]{Everaers2004}%
  \BibitemOpen
  \bibfield  {author} {\bibinfo {author} {\bibfnamefont {R.}~\bibnamefont
  {Everaers}}, \bibinfo {author} {\bibfnamefont {S.~K.}\ \bibnamefont
  {Sukumaran}}, \bibinfo {author} {\bibfnamefont {G.~S.}\ \bibnamefont
  {Grest}}, \bibinfo {author} {\bibfnamefont {C.}~\bibnamefont {Svaneborg}},
  \bibinfo {author} {\bibfnamefont {A.}~\bibnamefont {Sivasubramanian}}, \ and\
  \bibinfo {author} {\bibfnamefont {K.}~\bibnamefont {Kremer}},\ }\href@noop {}
  {\bibfield  {journal} {\bibinfo  {journal} {Science}\ }\textbf {\bibinfo
  {volume} {303}},\ \bibinfo {pages} {823} (\bibinfo {year}
  {2004})}\BibitemShut {NoStop}%
\bibitem [{\citenamefont {Svaneborg}\ and\ \citenamefont
  {Everaers}(2018)}]{Svaneborg2018a}%
  \BibitemOpen
  \bibfield  {author} {\bibinfo {author} {\bibfnamefont {C.}~\bibnamefont
  {Svaneborg}}\ and\ \bibinfo {author} {\bibfnamefont {R.}~\bibnamefont
  {Everaers}},\ }\href@noop {} {\bibfield  {journal} {\bibinfo  {journal}
  {arXiv:1808.03503}\ } (\bibinfo {year} {2018})}\BibitemShut {NoStop}%
\bibitem [{\citenamefont {Svaneborg}\ \emph {et~al.}(2018)\citenamefont
  {Svaneborg}, \citenamefont {Karimi-Varzaneh}, \citenamefont {Hojdis},
  \citenamefont {Fleck},\ and\ \citenamefont {Everaers}}]{Svaneborg2018b}%
  \BibitemOpen
  \bibfield  {author} {\bibinfo {author} {\bibfnamefont {C.}~\bibnamefont
  {Svaneborg}}, \bibinfo {author} {\bibfnamefont {H.~A.}\ \bibnamefont
  {Karimi-Varzaneh}}, \bibinfo {author} {\bibfnamefont {N.}~\bibnamefont
  {Hojdis}}, \bibinfo {author} {\bibfnamefont {F.}~\bibnamefont {Fleck}}, \
  and\ \bibinfo {author} {\bibfnamefont {R.}~\bibnamefont {Everaers}},\
  }\href@noop {} {\bibfield  {journal} {\bibinfo  {journal} {arXiv:1808.03509}\
  } (\bibinfo {year} {2018})}\BibitemShut {NoStop}%
\bibitem [{\citenamefont {Grest}(2016)}]{Grest2016}%
  \BibitemOpen
  \bibfield  {author} {\bibinfo {author} {\bibfnamefont {G.~S.}\ \bibnamefont
  {Grest}},\ }\href@noop {} {\bibfield  {journal} {\bibinfo  {journal} {J.
  Chem. Phys.}\ }\textbf {\bibinfo {volume} {145}},\ \bibinfo {pages} {141101}
  (\bibinfo {year} {2016})}\BibitemShut {NoStop}%
\bibitem [{\citenamefont {Wind}\ \emph {et~al.}(2003)\citenamefont {Wind},
  \citenamefont {Graf}, \citenamefont {Heuer},\ and\ \citenamefont
  {Spiess}}]{Wind2003}%
  \BibitemOpen
  \bibfield  {author} {\bibinfo {author} {\bibfnamefont {M.}~\bibnamefont
  {Wind}}, \bibinfo {author} {\bibfnamefont {R.}~\bibnamefont {Graf}}, \bibinfo
  {author} {\bibfnamefont {A.}~\bibnamefont {Heuer}}, \ and\ \bibinfo {author}
  {\bibfnamefont {H.~W.}\ \bibnamefont {Spiess}},\ }\href@noop {} {\bibfield
  {journal} {\bibinfo  {journal} {Phys. Rev. Lett.}\ }\textbf {\bibinfo
  {volume} {91}},\ \bibinfo {pages} {155702} (\bibinfo {year}
  {2003})}\BibitemShut {NoStop}%
\bibitem [{\citenamefont {Fetters}\ \emph {et~al.}(2007)\citenamefont
  {Fetters}, \citenamefont {Lohse},\ and\ \citenamefont {Colby}}]{Fetters2007}%
  \BibitemOpen
  \bibfield  {author} {\bibinfo {author} {\bibfnamefont {L.~J.}\ \bibnamefont
  {Fetters}}, \bibinfo {author} {\bibfnamefont {D.~J.}\ \bibnamefont {Lohse}},
  \ and\ \bibinfo {author} {\bibfnamefont {R.~H.}\ \bibnamefont {Colby}},\ }in\
  \href@noop {} {\emph {\bibinfo {booktitle} {Physical Properties of Polymers
  Handbook, 2nd}}},\ \bibinfo {editor} {edited by\ \bibinfo {editor}
  {\bibfnamefont {J.~E.}\ \bibnamefont {Mark}}}\ (\bibinfo  {publisher}
  {Springer},\ \bibinfo {address} {New York},\ \bibinfo {year} {2007})\
  Chap.~\bibinfo {chapter} {25}, pp.\ \bibinfo {pages} {447--454}\BibitemShut
  {NoStop}%
\bibitem [{\citenamefont {Zhang}\ \emph {et~al.}(2014)\citenamefont {Zhang},
  \citenamefont {Moreira}, \citenamefont {Stuehn}, \citenamefont {Daoulas},\
  and\ \citenamefont {Kremer}}]{Zhang2014}%
  \BibitemOpen
  \bibfield  {author} {\bibinfo {author} {\bibfnamefont {G.}~\bibnamefont
  {Zhang}}, \bibinfo {author} {\bibfnamefont {L.~A.}\ \bibnamefont {Moreira}},
  \bibinfo {author} {\bibfnamefont {T.}~\bibnamefont {Stuehn}}, \bibinfo
  {author} {\bibfnamefont {K.~C.}\ \bibnamefont {Daoulas}}, \ and\ \bibinfo
  {author} {\bibfnamefont {K.}~\bibnamefont {Kremer}},\ }\href@noop {}
  {\bibfield  {journal} {\bibinfo  {journal} {ACS Macro Lett.}\ }\textbf
  {\bibinfo {volume} {3}},\ \bibinfo {pages} {198} (\bibinfo {year}
  {2014})}\BibitemShut {NoStop}%
\bibitem [{\citenamefont {Moreira}\ \emph {et~al.}(2015)\citenamefont
  {Moreira}, \citenamefont {Zhang}, \citenamefont {M{\"u}ller}, \citenamefont
  {Stuehn},\ and\ \citenamefont {Kremer}}]{Moreira2015}%
  \BibitemOpen
  \bibfield  {author} {\bibinfo {author} {\bibfnamefont {L.~A.}\ \bibnamefont
  {Moreira}}, \bibinfo {author} {\bibfnamefont {G.}~\bibnamefont {Zhang}},
  \bibinfo {author} {\bibfnamefont {F.}~\bibnamefont {M{\"u}ller}}, \bibinfo
  {author} {\bibfnamefont {T.}~\bibnamefont {Stuehn}}, \ and\ \bibinfo {author}
  {\bibfnamefont {K.}~\bibnamefont {Kremer}},\ }\href@noop {} {\bibfield
  {journal} {\bibinfo  {journal} {Macromol. Theor. Simul.}\ }\textbf {\bibinfo
  {volume} {24}},\ \bibinfo {pages} {419} (\bibinfo {year} {2015})}\BibitemShut
  {NoStop}%
\bibitem [{\citenamefont {Hsu}\ and\ \citenamefont {Kremer}(2016)}]{Hsu2016}%
  \BibitemOpen
  \bibfield  {author} {\bibinfo {author} {\bibfnamefont {H.-P.}\ \bibnamefont
  {Hsu}}\ and\ \bibinfo {author} {\bibfnamefont {K.}~\bibnamefont {Kremer}},\
  }\href@noop {} {\bibfield  {journal} {\bibinfo  {journal} {J. Chem. Phys.}\
  }\textbf {\bibinfo {volume} {144}},\ \bibinfo {pages} {154907} (\bibinfo
  {year} {2016})}\BibitemShut {NoStop}%
\bibitem [{\citenamefont {Hsu}\ and\ \citenamefont
  {Kremer}(2018{\natexlab{a}})}]{Hsu2018a}%
  \BibitemOpen
  \bibfield  {author} {\bibinfo {author} {\bibfnamefont {H.-P.}\ \bibnamefont
  {Hsu}}\ and\ \bibinfo {author} {\bibfnamefont {K.}~\bibnamefont {Kremer}},\
  }\href@noop {} {\bibfield  {journal} {\bibinfo  {journal} {ACS Macro Lett.}\
  }\textbf {\bibinfo {volume} {7}},\ \bibinfo {pages} {107} (\bibinfo {year}
  {2018}{\natexlab{a}})}\BibitemShut {NoStop}%
\bibitem [{\citenamefont {Hsu}\ and\ \citenamefont
  {Kremer}(2018{\natexlab{b}})}]{Hsu2018b}%
  \BibitemOpen
  \bibfield  {author} {\bibinfo {author} {\bibfnamefont {H.-P.}\ \bibnamefont
  {Hsu}}\ and\ \bibinfo {author} {\bibfnamefont {K.}~\bibnamefont {Kremer}},\
  }\href@noop {} {\bibfield  {journal} {\bibinfo  {journal} {Phys. Rev. Lett.}\
  }\textbf {\bibinfo {volume} {121}},\ \bibinfo {pages} {167801} (\bibinfo
  {year} {2018}{\natexlab{b}})}\BibitemShut {NoStop}%
\bibitem [{\citenamefont {Rubinstein}\ and\ \citenamefont
  {Colby}(2003)}]{Rubinstein2003}%
  \BibitemOpen
  \bibfield  {author} {\bibinfo {author} {\bibfnamefont {M.}~\bibnamefont
  {Rubinstein}}\ and\ \bibinfo {author} {\bibfnamefont {R.~H.}\ \bibnamefont
  {Colby}},\ }\href@noop {} {\emph {\bibinfo {title} {Polymer Physics}}}\
  (\bibinfo  {publisher} {Oxford University Press, Oxford},\ \bibinfo {year}
  {2003})\BibitemShut {NoStop}%
\bibitem [{\citenamefont {Martyna}\ \emph {et~al.}(1994)\citenamefont
  {Martyna}, \citenamefont {Tobias},\ and\ \citenamefont
  {Klein}}]{Martyna1994}%
  \BibitemOpen
  \bibfield  {author} {\bibinfo {author} {\bibfnamefont {G.~J.}\ \bibnamefont
  {Martyna}}, \bibinfo {author} {\bibfnamefont {D.~J.}\ \bibnamefont {Tobias}},
  \ and\ \bibinfo {author} {\bibfnamefont {M.~L.}\ \bibnamefont {Klein}},\
  }\href@noop {} {\bibfield  {journal} {\bibinfo  {journal} {J. Chem. Phys.}\
  }\textbf {\bibinfo {volume} {101}},\ \bibinfo {pages} {4177} (\bibinfo {year}
  {1994})}\BibitemShut {NoStop}%
\bibitem [{\citenamefont {Quigley}\ and\ \citenamefont
  {Probert}(2004)}]{Quigley2004}%
  \BibitemOpen
  \bibfield  {author} {\bibinfo {author} {\bibfnamefont {D.}~\bibnamefont
  {Quigley}}\ and\ \bibinfo {author} {\bibfnamefont {M.~I.~J.}\ \bibnamefont
  {Probert}},\ }\href@noop {} {\bibfield  {journal} {\bibinfo  {journal} {J.
  Chem. Phys.}\ }\textbf {\bibinfo {volume} {120}},\ \bibinfo {pages} {11432}
  (\bibinfo {year} {2004})}\BibitemShut {NoStop}%
\bibitem [{\citenamefont {Halverson}\ \emph {et~al.}(2013)\citenamefont
  {Halverson}, \citenamefont {Brandes}, \citenamefont {Lenz}, \citenamefont
  {Arnold}, \citenamefont {Bevc}, \citenamefont {Starchenko}, \citenamefont
  {Kremer}, \citenamefont {Stuehn},\ and\ \citenamefont {Reith}}]{Espressopp}%
  \BibitemOpen
  \bibfield  {author} {\bibinfo {author} {\bibfnamefont {J.~D.}\ \bibnamefont
  {Halverson}}, \bibinfo {author} {\bibfnamefont {T.}~\bibnamefont {Brandes}},
  \bibinfo {author} {\bibfnamefont {O.}~\bibnamefont {Lenz}}, \bibinfo {author}
  {\bibfnamefont {A.}~\bibnamefont {Arnold}}, \bibinfo {author} {\bibfnamefont
  {S.}~\bibnamefont {Bevc}}, \bibinfo {author} {\bibfnamefont {V.}~\bibnamefont
  {Starchenko}}, \bibinfo {author} {\bibfnamefont {K.}~\bibnamefont {Kremer}},
  \bibinfo {author} {\bibfnamefont {T.}~\bibnamefont {Stuehn}}, \ and\ \bibinfo
  {author} {\bibfnamefont {D.}~\bibnamefont {Reith}},\ }\href@noop {}
  {\bibfield  {journal} {\bibinfo  {journal} {Comput. Phys. Commun.}\ }\textbf
  {\bibinfo {volume} {184}},\ \bibinfo {pages} {1129} (\bibinfo {year}
  {2013})}\BibitemShut {NoStop}%
\bibitem [{\citenamefont {Rouse}(1953)}]{Rouse1953}%
  \BibitemOpen
  \bibfield  {author} {\bibinfo {author} {\bibfnamefont {P.~R.}\ \bibnamefont
  {Rouse}},\ }\href@noop {} {\bibfield  {journal} {\bibinfo  {journal} {J.
  Chem. Phys.}\ }\textbf {\bibinfo {volume} {21}},\ \bibinfo {pages} {1272}
  (\bibinfo {year} {1953})}\BibitemShut {NoStop}%
\bibitem [{\citenamefont {Vogel}(1921)}]{Vogel1921}%
  \BibitemOpen
  \bibfield  {author} {\bibinfo {author} {\bibfnamefont {H.}~\bibnamefont
  {Vogel}},\ }\href@noop {} {\bibfield  {journal} {\bibinfo  {journal} {Phys.
  Z}\ }\textbf {\bibinfo {volume} {22}},\ \bibinfo {pages} {645} (\bibinfo
  {year} {1921})}\BibitemShut {NoStop}%
\bibitem [{\citenamefont {Fulcher}(1925)}]{Fulcher1925}%
  \BibitemOpen
  \bibfield  {author} {\bibinfo {author} {\bibfnamefont {G.~S.}\ \bibnamefont
  {Fulcher}},\ }\href@noop {} {\bibfield  {journal} {\bibinfo  {journal} {J.
  Am. Ceram. Soc.}\ }\textbf {\bibinfo {volume} {8}},\ \bibinfo {pages} {339}
  (\bibinfo {year} {1925})}\BibitemShut {NoStop}%
\bibitem [{\citenamefont {Tammann}\ \emph {et~al.}(1926)\citenamefont
  {Tammann}, \citenamefont {Hesse},\ and\ \citenamefont {Anorg}}]{Tammann1926}%
  \BibitemOpen
  \bibfield  {author} {\bibinfo {author} {\bibfnamefont {G.}~\bibnamefont
  {Tammann}}, \bibinfo {author} {\bibfnamefont {W.}~\bibnamefont {Hesse}}, \
  and\ \bibinfo {author} {\bibfnamefont {Z.}~\bibnamefont {Anorg}},\
  }\href@noop {} {\bibfield  {journal} {\bibinfo  {journal} {Allgem. Chem.}\
  }\textbf {\bibinfo {volume} {156}},\ \bibinfo {pages} {245} (\bibinfo {year}
  {1926})}\BibitemShut {NoStop}%
\bibitem [{\citenamefont {Nascimento}\ and\ \citenamefont
  {Aparicio}(2007)}]{Nascimento2007}%
  \BibitemOpen
  \bibfield  {author} {\bibinfo {author} {\bibfnamefont {M.~L.~F.}\
  \bibnamefont {Nascimento}}\ and\ \bibinfo {author} {\bibfnamefont
  {C.}~\bibnamefont {Aparicio}},\ }\href@noop {} {\bibfield  {journal}
  {\bibinfo  {journal} {J. Phys. Chem. Solids}\ }\textbf {\bibinfo {volume}
  {68}},\ \bibinfo {pages} {104} (\bibinfo {year} {2007})}\BibitemShut
  {NoStop}%
\end{thebibliography}

\end{document}